\documentclass[usenatbib]{mn2e}
\usepackage{amssymb}
\usepackage{graphicx}
\usepackage{longtable}

\def\Teff{$T_{\rm eff}$}
\def\logg{$\log\,g$}

\def\Vt{V${\rm t}$}

\def\aap{A\&A}%
\newcommand {\apgt} {\ {\raise-.5ex\hbox{$\buildrel>\over\sim$}}\ }
\newcommand {\aplt} {\ {\raise-.5ex\hbox{$\buildrel<\over\sim$}}\ }

\title[Gd, Dy, and Th]
{Enrichment of the Galactic disc with neutron-capture elements: Gd, Dy, and Th
}
\author[T.~Mishenina  et al.]
{T.~Mishenina$^{1}$ \thanks{tmishenina@ukr.net;* mpignatari@gmail.com},
 M.~Pignatari$^{2,3,4,5}$ \thanks{The NuGrid collaboration, http://www.nugridstars.org} *,
T.~Gorbaneva$^{1}$,  
B.~C\^{o}t\'e$^{9,5,2}$ $\dagger$ , \newauthor
A.~Yag\"ue L\'opez$^{2,10}$ $\dagger$,
F.-K.~Thielemann$^{6,7}$,
 C.~Soubiran$^{8}$  
 \\
$^{1}$Astronomical Observatory, Odessa National University, Shevchenko Park, 65014, Odessa, Ukraine\\
$^{2}$ Konkoly Observatory, Research Centre for Astronomy and Earth Sciences (CSFK), E\"otv\"os Lor\'and Research Network (ELKH),\\ 
Konkoly Thege Mikl\'{o}s \'{u}t 15-17, H-1121 Budapest, Hungary\\
$^{3}$ CSFK, MTA Centre of Excellence, Budapest, Konkoly Thege Mikl\'{o}s \'{u}t 15-17, H-1121, Hungary\\
$^{4}$  E.A. Milne Centre for Astrophysics, Dept of Physics \& Mathematics, University of Hull, HU6 7RX, United Kingdom  \\
$^{5}$ Joint Institute for Nuclear Astrophysics - Center for the Evolution of the Elements, USA\\
$^{6}$ Department of Physics, University of Basel, Klingelbergstrabe 82,
        4056 Basel, Switzerland\\
$^{7}$ GSI Helmholtzzentrum für Schwerionenforschung, Planckstrasse 1, D-64291 Darmstadt, Germany\\
$^{8}$  Laboratoire d'Astrophysique de Bordeaux, 
        Univ. Bordeaux  - CNRS, B18N,  all\'ee Geoffroy Saint-Hilaire, 33615 Pessac, France\\
$^{9}$ Department of Physics and Astronomy, University of Victoria, Victoria, BC V8P 5C2, Canada\\
$^{10}$ Computer, Computational and Statistical Sciences (CCS) Division, Center for Theoretical Astrophysics, Los Alamos National\\ Laboratory, Los Alamos, NM 87545, USA\\
}

\begin{document}

\date{Accepted 2021 xxx. Received 2021 xxx; in original form 2021 xxx}
\pagerange{\pageref{firstpage}--\pageref{lastpage}}
\pubyear{2021}

\maketitle

\label{firstpage}

\begin{abstract}
The study of the origin of heavy elements is one of the main goals of nuclear astrophysics.
In this paper, we present 
new observational data  
for the heavy $r$-process elements gadolinium (Gd, Z=64), dysprosium (Dy, Z=66) and thorium (Th, Z=90) in a sample of 276 Galactic disc stars ( --1.0$<$[Fe/H]$<$+0.3). The stellar spectra  
have a high resolution of 42,000 and 75,000, and the signal-to-noise ratio 
higher than 100. The LTE abundances of Gd, Dy and Th have been determined by comparing the observed and synthetic spectra for three Gd lines (149 stars), four Dy lines (152 stars) and the Th line at 4019.13 \AA~(170 stars). For about 70\% of the stars in our sample Gd and Dy are measured for the first time, and Th for 95\% of the stars. 
Typical errors vary from 0.07 to 0.16 dex. This paper provides the first extended set of Th observations in the Milky Way disc.
Together with 
europium (Eu, Z = 63) data from our previous studies, we have compared these new observations with nucleosynthesis predictions and Galactic Chemical Evolution simulations.
We confirm that [Gd/Fe] and [Dy/Fe] show the same behavior of Eu. We study with GCE simulations the evolution of [Th/Fe] in comparison with [Eu/Fe], showing that unlike Eu either the Th production is metallicity dependent in case of a unique source of the r-process in the Galaxy, or the frequency of the Th-rich r-process source is decreasing with the increasing of [Fe/H]. 
\end{abstract}


\section{Introduction}
The nucleosynthesis of heavy neutron-capture elements in stars and their observations is one of the main research drivers for modern nuclear astrophysics. In this context, the origin of the rapid neutron capture process \citep[r-process, e.g.,][and references therein]{cowan:21} is still a major matter of debate. 
Among others, the most favoured $r$-process sites are neutron-star mergers \cite[e.g.][]{eichler:89, freiburghaus:99, goriely:15, thielemann:17,rosswog:18} and neutron star-black hole mergers \cite[e.g.][]{lattimer:74, surman:08,Fernandez.Foucart.Lippuner:2020}, certain rare classes of fast-rotating supernovae with powerful magnetic fields \cite[e.g.][]{symbalisty:85, cameron:03,nishimura:06, winteler:12,nishimura:2017,moesta:18,obergaulinger:18,reichert:21}, as well as hypernovae or collapsars \cite[e.g.][]{cameron:03,siegel:19,thielemann:20, zenati:20, brauer:21}. In the past, core-collapse supernovae (CCSNe) have been considered as the dominant source of the $r$-process, initially by suggesting neutron-rich innermost ejecta \cite[e.g.][]{hillebrandt:76}, later arguing for fast $(\alpha,n)$-reactions in explosive burning of He shells \cite[e.g.][]{truran:78,thielemann:79,cowan:85}, and afterwards turning to high entropy conditions in neutrino-driven winds during the core-collapse and explosion phase \cite[e.g.][]{woosley:94, takahashi:94,hoffman:97,ning:07, farouqi:10, arcones:13}. At present, realistic CCSN simulations do not provide the right conditions to produce a complete $r$-process pattern, while still a mild weak $r$-process production could be possible \cite[see e.g.,][]{wanajo:11,curtis:19,cowan:21,ghosh:21}.

Stellar spectroscopic observations can be used to derive fundamental constraints for theoretical simulations. In particular, a large number of works in the past decade has been made to define the composition of old $r$-process rich stars, formed in the early Milky Way Galaxy
\cite[e.g.][]{sneden:03, simmerer:04, beers:05, barklem:05, yong:13, roederer:14, sakari:18, mashonkina:14a, hansen:18}. 
Such an importance is primarily 
due to the possibility to trace contributions of one or several stellar sites of production of these elements within the early Galaxy timescales, before 
global gas mixing might actually take place  \cite[e.g.][]{hansen:20}. Therefore, stellar observations can be used to test directly 
$r$-process predictions from different stellar sites 
\citep{farouqi:21} – for instance, those resulting from neutron-star mergers \citep{ji:19} or from magnetorotational hypernovae \cite[][]{yong:21}.
This includes to study the role of progenitors of satellite galaxies on the early galactic chemical enrichment \cite[e.g.][]{gudin:21}.

As the Galaxy evolves, new stars are forming enriched by previous stellar generations. Compared to the early Galaxy, different stellar sources need to be taken into account for the production of neutron-capture elements during the chemical evolution of the Galaxy \citep[GCE, e.g.,][]{prantzos:18, kobayashi:20}.
At present in the Milky Way there are two main processes responsible for the production of 
heavy elements. In addition to the $r$-process, the slow neutron capture process \citep[s-process, e.g.,][]{kaeppeler:11} is responsible for about half of the abundances beyond iron in the solar system. 
The s-process elements are mainly produced in massive stars \cite[e.g.][]{the:07, pignatari:10, frischknecht:16, limongi:18} and in Asymptotic Giant Branch (AGB) stars \cite[e.g.][]{gallino:98, busso:99, bisterzo:14, cristallo:15, karakas:16, battino:19}. 
In order to take into account the $r$-process contribution in GCE calculations, these yields are often derived from the solar residual method: the $r$-process abundance pattern is obtained from the solar composition after removing the $s$-process contribution, and then it is assumed to be the same for all metallicities \citep[e.g.,][]{travaglio:04, prantzos:18}. Alternatively, a large range of theoretical $r$-process yields may be adopted. 
GCE models and simulations are crucial tools to better understand the evolution of $r$-process elements in galaxies (e.g., \citealt{wehmeyer:15,naiman:18,vandeVoort:20}). In particular, the [Eu/Fe] vs [Fe/H] trend in the Galactic disc has been targeted several times to probe the enrichment timescales and contribution of neutron star mergers and rare classes of core-collapse supernovae. Studies have suggested that neutron star mergers alone cannot reproduce the decreasing trend of Eu when assuming a merger delay-time probability distribution (DTD) in the form of $t^{-1}$ (e.g., \citealt{cote:17,cote:19,hotokezaka:18,haynes:19,simonetti:19}). Such an issue, however, can be lifted by implying metallicity-dependent DTDs (e.g., \citealt{simonetti:19}), imposing shorter delay times for mergers relative to Type~Ia supernovae (e.g., \citealt{matteucci:14,wehmeyer:15,cote:17,cavallo:21,wanajo:21}), or adopting different treatments for how $r$-process elements are mixed and distributed within the Galaxy (e.g., \citealt{schonrich:19,banerjee:20,beniamini:20}). Another solution to recover the [Eu/Fe] trend is to involve additional sources alongside neutron star mergers for Eu, such as rare supernovae originating from massive stars (e.g., \citealt{cote:19,siegel:19,kobayashi:20,cavallo:21,greggio:21,farouqi:21}).
 
Europium is the most extensively studied chemical element produced via the $r$-process in the Galactic disc (according to \citep[e.g.][]{bisterzo:14}, the solar $s$-process contribution is about 6 \%). Europium abundance in the Galactic disc has been investigated by many researchers \cite[e.g.][etc.]{mashonkina:01, reddy:03, bensby:05, mishenina:13}. On the other hand, there are only a limited amount of stars with available several $r$-process elements measured together from the same analysis. 
For instance, in \cite{guiglion:18} 
gadolinium and dysprosium were examined together with europium and barium 
within the frameworks of the AMBRE Project based on high-resolution FEROS, HARPS and UVES spectra from the ESO archive. The contribution by the $s$-process to the Gd and Dy solar abundances is estimated 
to be 15.4 \% and 15.0 \%, respectively \citep[][]{bisterzo:14}, i.e., indicating a dominant $r$-process contribution. 
Gd, and Dy abundances in thin and thick discs were investigated by \cite{guiglion:18} and in solar twins by \cite{spina:18}. Th abundances and the ratios Th/Eu were obtained for thin disc stars to estimate the age of the disc \citep{peloso:05}. The Th abundance is also measured for samples of solar twins stars \citep{unterborn:15, botelho:19}.
As a continuation of our previous research focused on studying of the Galactic disc enrichment with neutron-capture elements \cite[][]{mishenina:13, mishenina:17, mishenina:19a, mishenina:19b}, this paper aims to investigate the abundance distribution of  
the $r$-process elements Gd, Dy and Th. 
Our study includes new Gd and Dy measurements for nearly 70 \% of the stars in our sample. For more than 90 \% of stars we present new Th values, and the GCE of the Milky Way disk is done for the first time taking into account both actinide (Th) and lanthanide (Eu, Gd and Dy) observations.

The paper is organized as follows. 
The observations and  the definition of the main stellar parameters are described in \S \ref{sec: stellar param}. 
The abundance determinations  and the error analysis are presented in \S \ref{sec: abundance determination}. 
The analysis of the behaviour of elemental abundances in the pattern of the theory of nucleosynthesis and the chemical evolution of the Galaxy is reported in \S \ref{sec: result, gce}. 
Conclusions are drawn in \S \ref{sec: conclusions}.

\section{Observations and atmospheric parameters}
\label{sec: stellar param}

The present study was carried out 
on an initial 
list of 276 stars and based on the spectra and atmospheric parameters 
by \cite{mishenina:13}.
The 1.93-m telescope at the Observatoire de Haute-Provence (OHP, France) and the echelle-type spectrograph ELODIE \cite[][]{baranne:96} were employed to obtain spectra at the resolving power R = 42,000 in the wavelength range from 4400 to 6800 \AA~and with the signal-to noise (S/N) better than 100 at 
5500 \AA. We also used additional spectra from the OHP spectroscopic archive \cite[][]{moultaka:04} collected with the SOPHIE spectrograph \cite[][]{perruchot:08} and covering a similar wavelength range at the spectral resolution R = 75,000.
The complex pre-processing of images available on-line and enabling to obtain spectroscopic data in a digital form was carried out immediately during observations \cite[][]{katz:98}. The subsequent processing of the studied spectra was performed using the DECH30 software package developed by G.A. Galazutdinov (see http://www.gazinur.com/DECH-software.html).
DECH software provides all stages of the CCD echelle spectral image processing, including bias/background subtraction, flat-field correction (separation), extraction of one-dimensional spectrum from two-dimensional images, diffuse light correction, spectrum addition and exclusion of cosmic-ray features. The programme enables to locate a fiducial continuum, to measure equivalent widths (EWs) of lines by several methods, to determine line positions and shifts and much more besides. In this case, we worked with spectra in the FITS format, using such options as normalisation of individual spectra to the local continuum, identification of spectral lines, development of the dispersion curve, measurements of the line depths and equivalent widths (EWs), elimination of cosmic-ray effects, selection of individual parts of the spectrum, etc. The measured line depths were subsequently used to determine the effective temperature (\Teff) while the EWs of the neutral and ionised iron lines were measured by the Gaussian profile fitting and employed to derive atmospheric parameters (the surface gravity, \logg, and micro-turbulent velocity, \Vt).

The stellar atmospheric parameters for the stars under examination in this work were determined by us in previous studies. The procedures employed to derive the effective temperatures \Teff, surface gravities \logg~ and microturbulent velocity \Vt~ for the target stars had been described in detail in \cite[][]{mishenina:01}  and \cite[][]{mishenina:04, mishenina:08}. In particular, the effective temperatures \Teff~ were determined by calibrating the line-depth ratios for the pairs of spectral lines that have different low-level excitation potentials with the application of the technique introduced and developed by \cite{kovtyukh:03}. For most of metal-poor stars in our sample, \Teff~ were assumed by adjusting the far-wings of the H$_\alpha$ line \citep{mishenina:01}. In \cite[][]{mishenina:04}, we showed that the temperature scales adopted in \cite[][]{mishenina:01, kovtyukh:03} are consistent.
The surface gravities, \logg, were computed from the ionisation equilibrium, which means that similar iron abundances should be obtained from the neutral iron (Fe I) and singly ionised iron (Fe II) lines. In our case, the difference between these values does not exceed 0.03 dex. The microturbulent velocity, \Vt, was derived by factoring out correlations between the iron abundances from Fe I lines and the equivalent widths (EW) of those Fe I lines. We adopted the iron abundance determined from the Fe I lines as the metallicity, [Fe/H].  As is known \cite[e.g][]{thevenin:99, shchukina:01, mashonkina:11, bergemann:12}, the lines of neutral iron are influenced by the deviations from the LTE in the solar and stellar spectra, and hence, these deviations also affect the iron abundances determined from those lines. However, within the temperature and metallicity ranges of our target stars, the NLTE corrections is less than 0.1 dex \cite[e.g][]{mashonkina:11}. Thus, both in the case of accepted [Fe/H] as the iron abundance from Fe I lines and also in the case of using the ionization equilibrium method for iron to derived \logg, this correction does not exceed the errors in determining of these parameters.  

The list of parameter values obtained, as well as their comparison with the results of other authors, has been given in \cite[][]{mishenina:04, mishenina:08, mishenina:13}. The estimated accuracy of our parameter determinations is as follows: $\Delta$(\Teff)= $\pm$100 K, $\Delta$(\logg)= $\pm$0.2dex, $\Delta$(\Vt) = $\pm$0.2km s$^{-1}$ and $\Delta$([Fe/H]) = $\pm$0.1dex. In this paper, we have compared our parameters with those obtained recently in the studies by \cite{guiglion:18} and \cite{spina:18}, wherein gadolinium and dysprosium abundances were determined, and also those in the studies by \cite{peloso:05}, who reported europium and thorium abundances (five stars in common with our sample) and \cite{unterborn:15} for one star in common with our sample; (see Table \ref{ncap}).
To compare with our findings, we chose the data reported by \cite[][]{guiglion:18} for the stars in common with the highest S/N among those available in on-line catalogs. We obtained average differences and errors for 36 stars in common by deducting our data from those by \cite{guiglion:18} (see Table \ref{ncap}). Then, we sorted out the data with \Teff~ from 5100 K to 6300 K and surface gravities within 3.5 $<$ \logg $<$ 5.0 (for 26 stars in common) as such ranges of parameter values had been chosen by the authors as criteria for the selection of stars for further analysis; the resulting differences are slightly smaller (Table \ref{ncap}). In general, we see a good agreement between our findings and those from the literature, as well as a good consistency with the estimated accuracy in parameter determinations adopted earlier. As we can see from Table \ref{ncap}, the  mean difference $\Delta$\Teff~ between our effective temperature and that obtained by other authors, does not exceed 25 K, and the rms deviations are within 100 K. A mean difference in gravity values $\Delta$\logg~ does not exceed 0.10, with the rms deviation is only slightly exceeding (0.22), adopted earlier (0.2). In terms of metallicity, the mean value does not exceed $\Delta$([Fe/H])= 0.05 $\pm$0.07 dex.

\begin{table*}
\begin{center}
\caption[]{Comparison of  parameters and  Eu, Gd, Dy, and Th abundance determinations taken from the literature with 
our results  for the $n$ stars common with our stellar sample. Our data for Eu abundances are from \cite[][]{mishenina:13}.
}
\label{ncap}
\begin{tabular}{ccccccccc}
\hline
 Reference & $\Delta$(\Teff) & $\Delta$(\logg) & $\Delta$([Fe/H]) &  $\Delta$([Eu/Fe]) & $\Delta$([Gd/Fe]) &$\Delta$([Dy/Fe]) &$\Delta$([Th/Fe]) & n \\
 \hline
Guiglion et al.&-14.7& 0.09 & 0.01 &  0.09 &0.09 &0.14  & -- &36\\
 2018 &$\pm$99.4 & $\pm$0.22 & $\pm$0.07 &  $\pm$0.16&$\pm$0.18 &$\pm$0.16 &--&  \\
Guiglion et al.& 12.1 & 0.06 & 0.0 & 0.09 &0.03  &0.10  & -- &  26 \\
 2018&$\pm$93 & $\pm$0.21 & $\pm$0.07 & $\pm$0.15 & $\pm$0.13 &$\pm$0.13& --&  \\
Spina et al.& 14.8 & 0.06 & 0.04 & --0.03 & --0.03 &--0.04 &-- & 6 (4)\\
 2018  &$\pm$25& $\pm$0.1 & $\pm$0.05 & $\pm$0.04&$\pm$0.04&$\pm$0.05& -- &  \\
del Peloso et al. & -16.4 & 0.04 & 0.05 & 0.05 &-- &-- & -0.23 & 5 (4)  \\
 2005 &$\pm$55 & $\pm$0.08 & $\pm$0.08 & $\pm$0.08 &-- &-- &$\pm$0.14 &  \\
Morell et al. & -3.8 & 0.02 & -0.02 & -- & -- &-- & -0.11 & 5 (5)  \\
 1992 &$\pm$85 & $\pm$0.20 & $\pm$0.04 & -- &  -- &-- &$\pm$0.12 &  \\
Unterborn et al.  & 23 & 0.05 & 0.04 & -- & --  &-- &0.48  &    1(1) \\
 2015  &-- & -- & -- &  -- & -- &  --& --& \\
Botelho et al. & -15.6 & -0.05 & -0.04 & 0.02 &-- &-- & -0.09 & 5(3)   \\
2019 &$\pm$28 & $\pm$0.11 & $\pm$0.05 & $\pm$0.04 &-- &-- & $\pm$0.15 &  \\
\hline
\end{tabular}
\end{center}
\end{table*}

We adopt the kinematic classification of the stars into the thin and thick discs and Hercules stream, as previously described in \cite[][]{mishenina:13}. TM: To determine the components of spatial velocity (U, V, W) and the belonging of stars to different galactic populations, the Hipparhos catalog was used.
Since the stars in our sample are bright and tend to have Gaia astrometric errors equivalent to those of the Hipparcos observations, we have not updated our classification with respect to the latest astrometric data from the Gaia Data Release 2 \citep{GDR2:18}. Some stars are even too bright to be measured by Gaia. Our previous sample of 276 stars in total consists of 21 stars belonging to the thick disc, 212 of those in the thin disc, 16 stars related to the Hercules stream and 27 unclassified stars.

\section{Abundance determination }
\label{sec: abundance determination}

\begin{figure}
\begin{tabular}{c}
\includegraphics[width=\columnwidth]{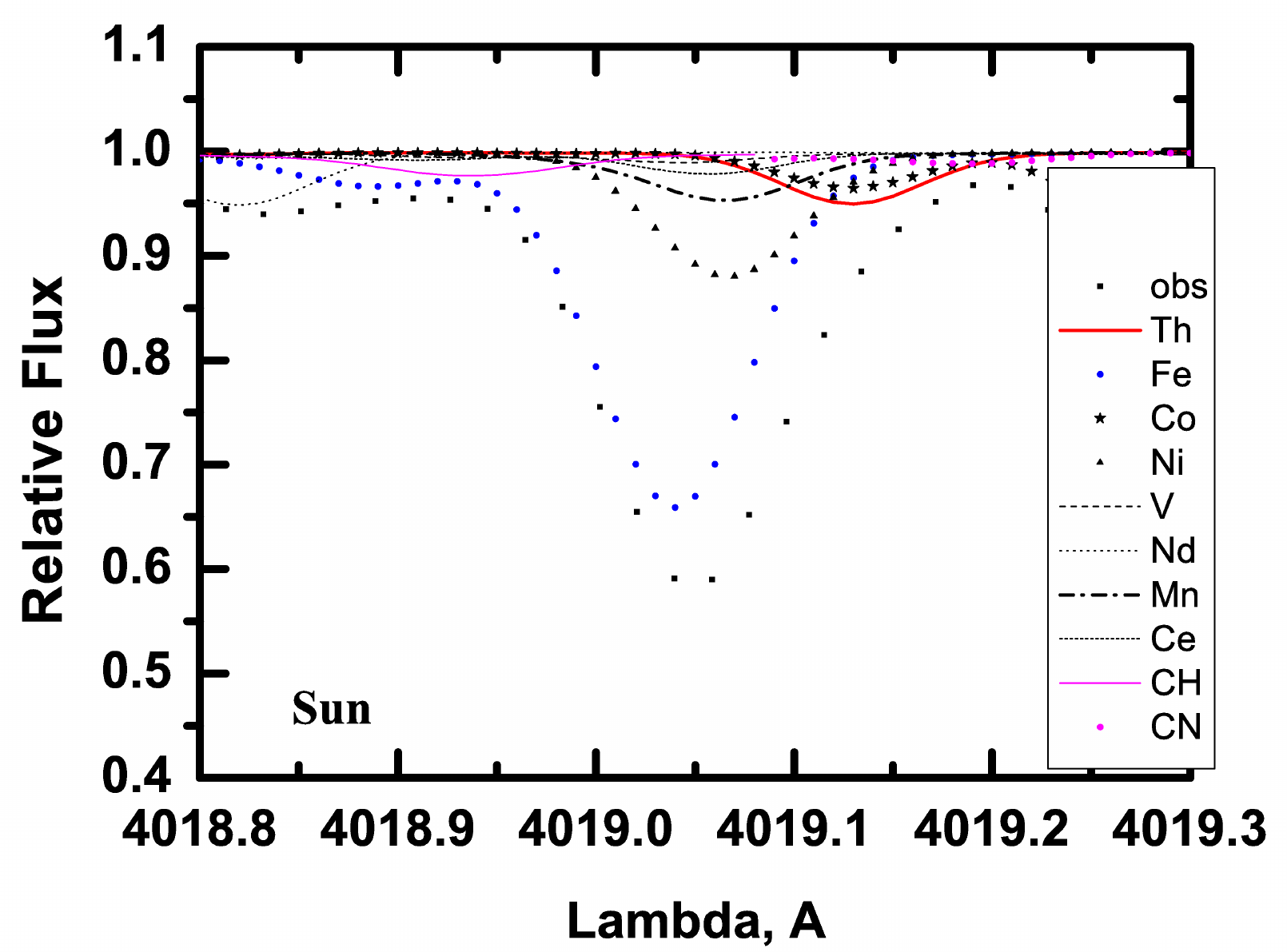}\\
\includegraphics[width=\columnwidth]{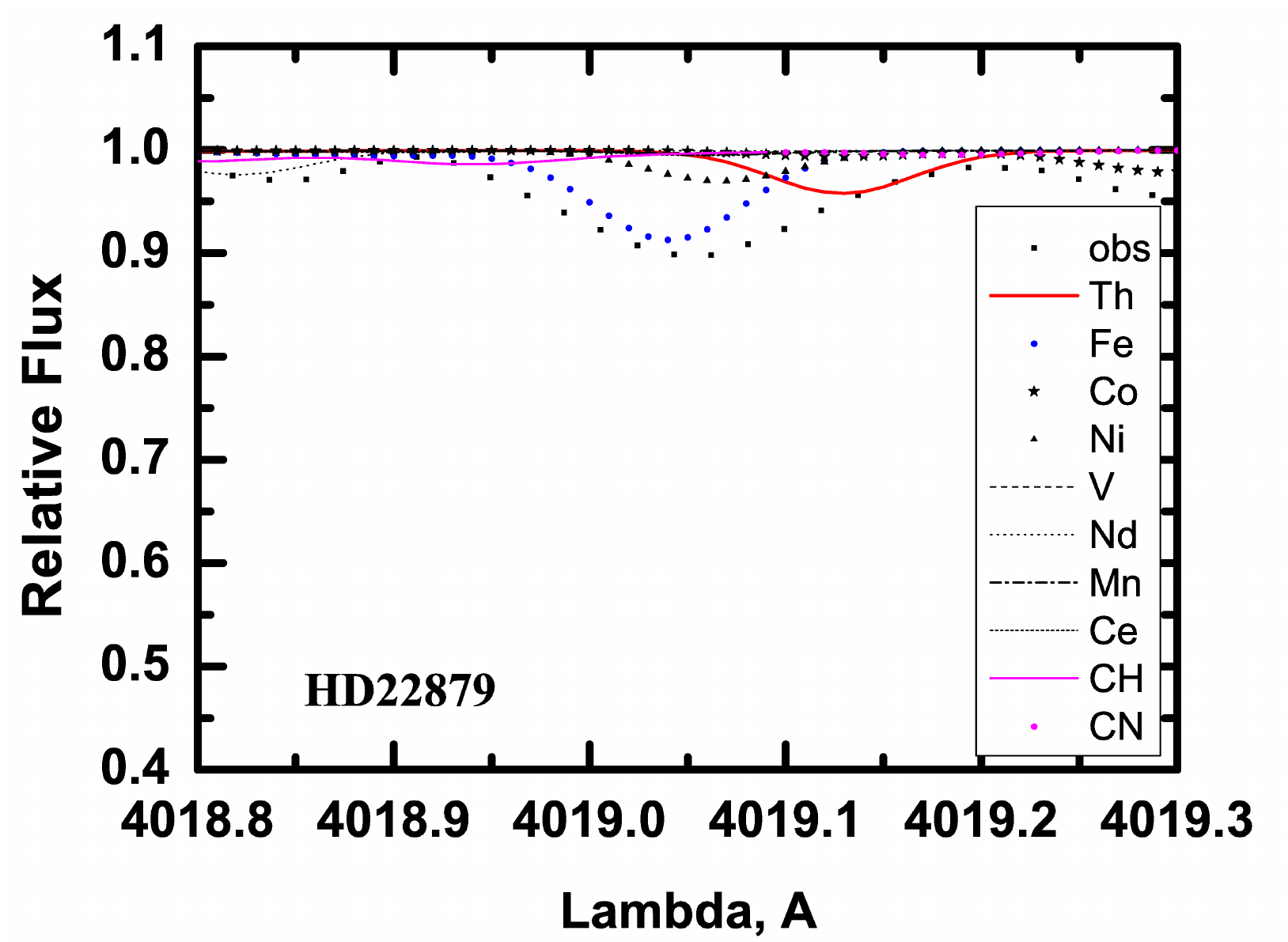}\\
\includegraphics[width=\columnwidth]{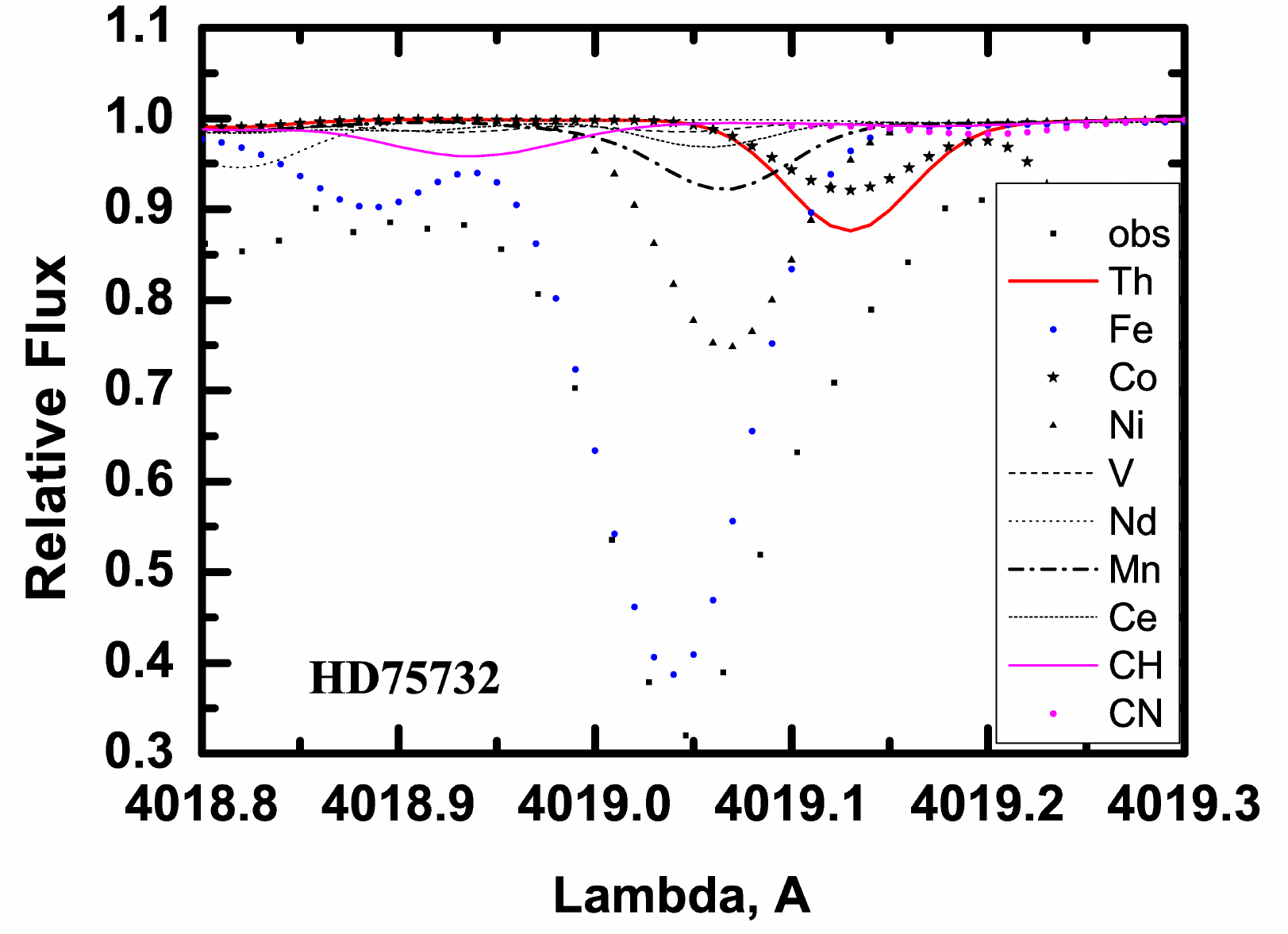}\\
\end{tabular}
\caption{Observed (squares) and calculated  spectra in the region of Th II line for Sun, HD 22879 (5825; 4.42; -0.91), and HD 75732 (5373; 4.30; 0.25). The contributions of various elements to the profile of the thorium line are marked on the panel. }
\label{th_sun_prof}
\end{figure}

\begin{figure}
\begin{tabular}{ccc}
\includegraphics[width=\columnwidth]{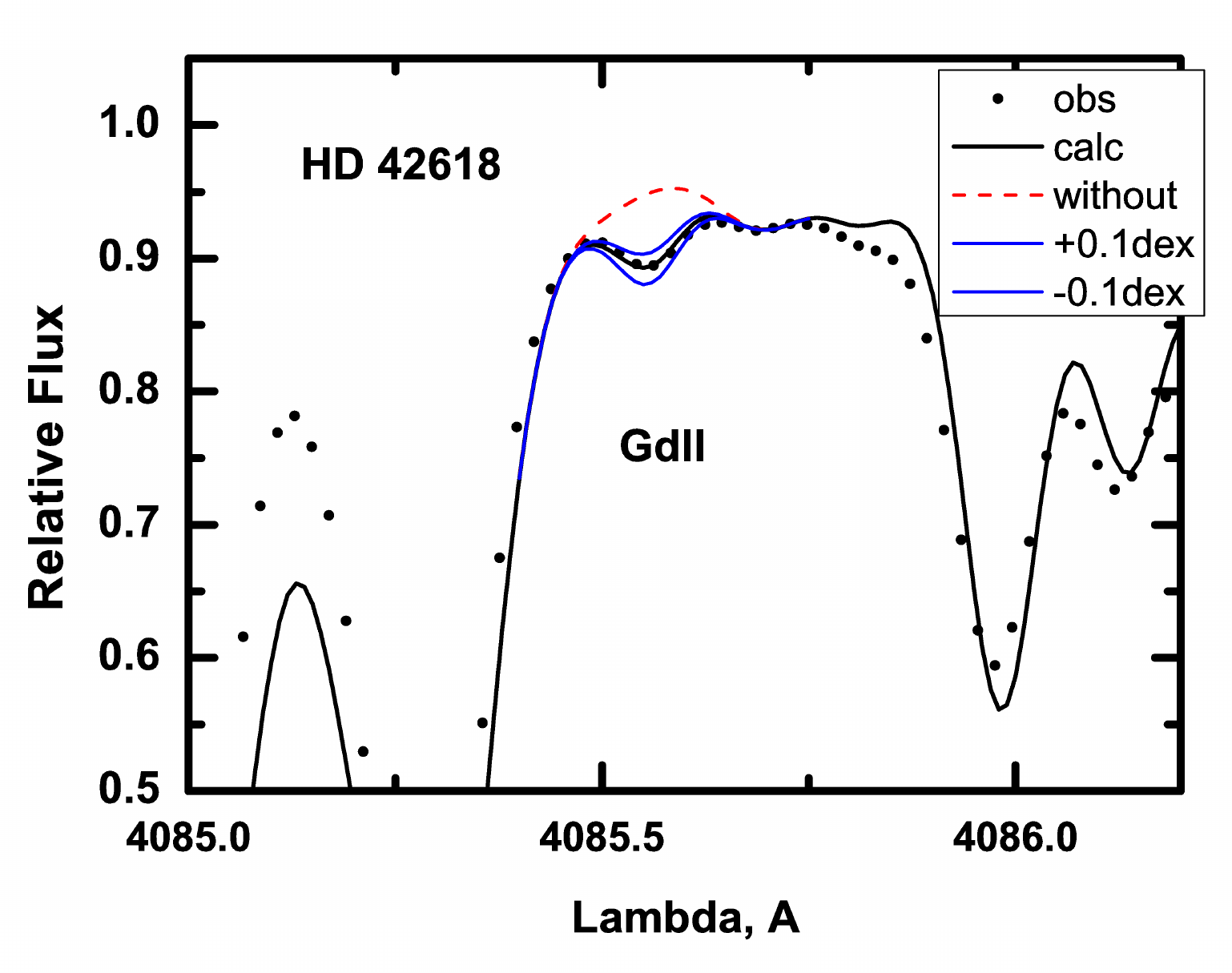}\\
\includegraphics[width=\columnwidth]{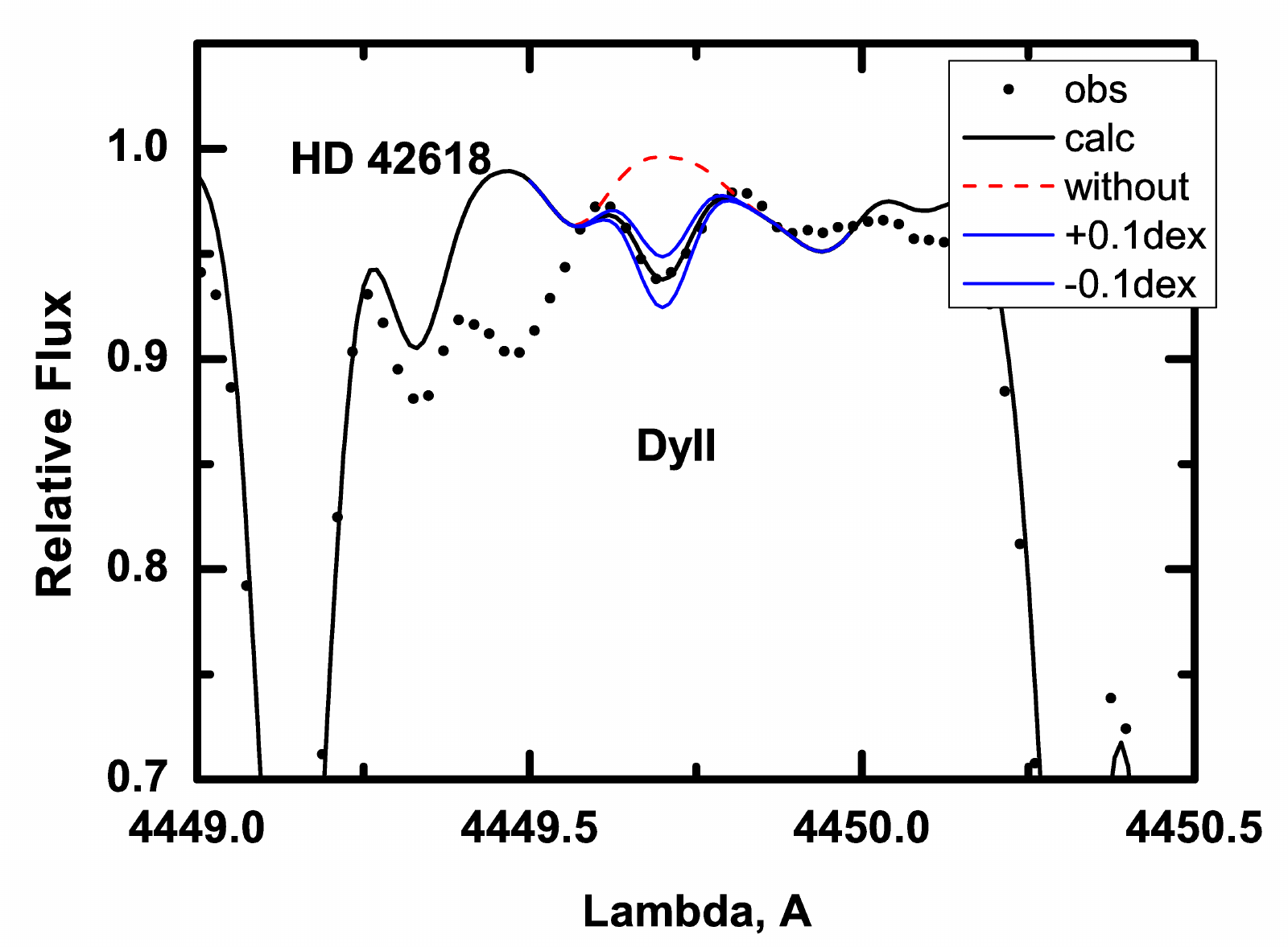}\\
\includegraphics[width=\columnwidth]{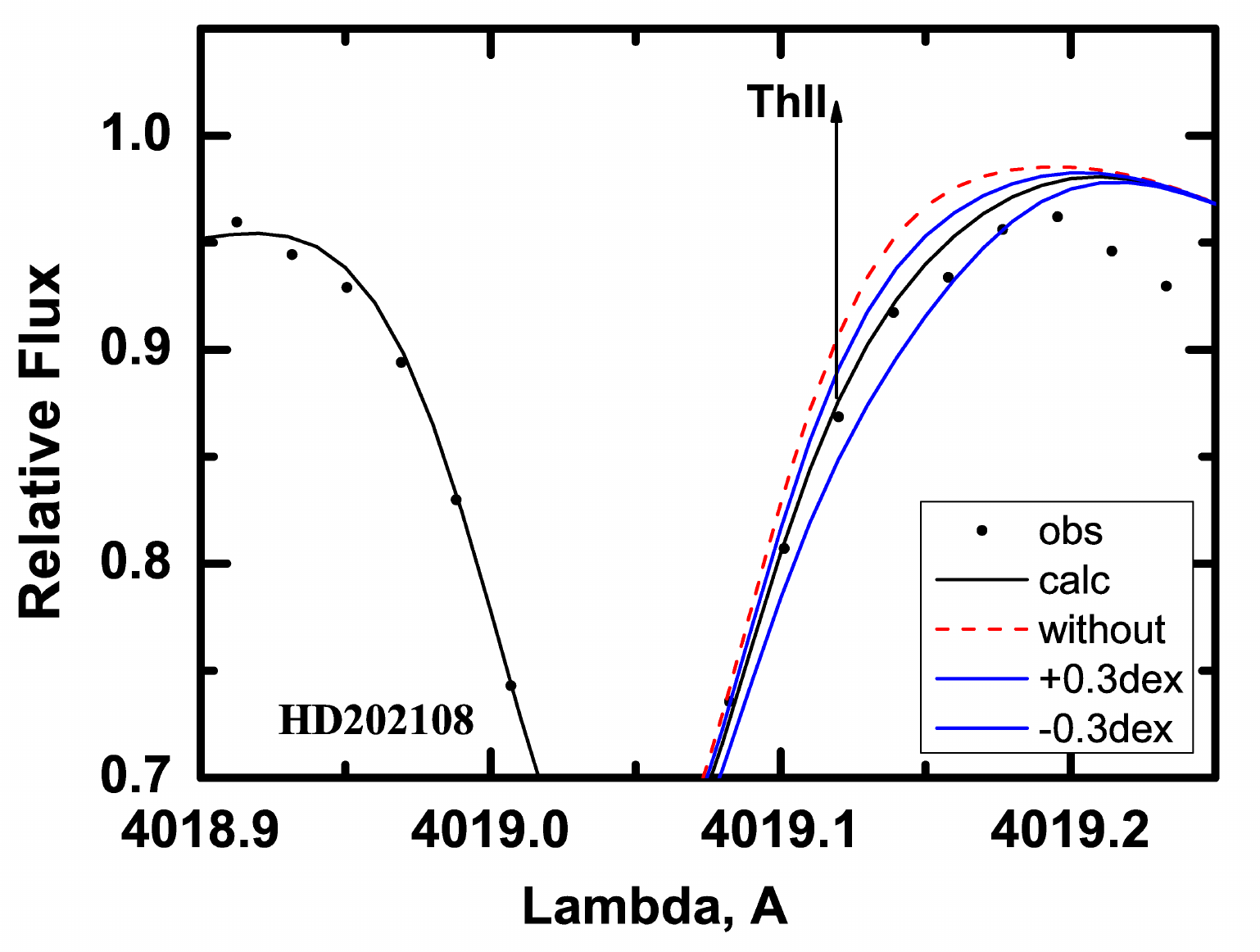}\\
\end{tabular}
\caption{Observed (points) and calculated spectra in the region of Gd II, Dy II and Th II lines for stars HD 42018 (5787;4.5;--0.07) and HD 202108 (5712;4.2;--0.21).}
\label{lines_prof}
\end{figure}

The abundances of Dy, Gd and Th were derived in the Local Thermodynamical Equilibrium (LTE) approximation with a new modified STARSP LTE spectral synthesis code \cite[][]{tsymbal:96} using the models by \cite{castelli:04}. For each star, the model was chosen by standard interpolation for \Teff~ and \logg. The metalicity [Fe/H] and the turbulent velocity \Vt~ are not interpolated,  in terms of metallicity, a model close to [Fe/H] of stars in $\pm$0.2 dex was selected and the turbulent velocity \Vt~ determined for each star was used.
 For Gd II lines 4037.89, 4085.56, 4483.33\AA, and Dy II lines 4073.12, 4077.97, 4103.31, and 4449.70 \AA, and Th II 4019.12 \AA~ the oscillator strengths log\,gf were adopted from last version (2016) of the VALD database \citep{kupka:99}.
In contrast to the considered Gd and Dy lines, the 4019.129 \AA~ Th line is a complex blend with a contribution to its intensity from the Th and Co abundances at almost the same wavelength \cite[e.g.][]{peloso:05, mashonkina:14b, botelho:19}.
 Using a list of VALD lines, which includes atomic and molecular lines, to describe the Th line in the solar spectrum, we found a noticeable discrepancy between the observed and calculated spectra in the region of Fe lines; to eliminate this, we corrected the log\,gf Fe I oscillator strengths with an appropriate fit. 
 The values of the oscillator strength adopted by us for the Th and Co lines follow the VALD list, namely, the values of log\,gf = -0.228 for the Th II line \citep[][]{nilsson:02} and log\,gf = -2.270 for the Co I line \citep[][]{lawler:90}, in our case, they are presented with a detailed contributions of the hyperfine structure.
  A list of our main atomic and molecular lines  employed in the thorium 4019 \AA~line region are given in Table \ref{list_thline}.  For the Sun, and two stars with stellar parameters (\Teff, \logg, [Fe/H]) HD 22879 (5825; 4.42; -0.91), and HD (5373; 4.30; 0.25) the predominant lines in the region are shown in Fig. \ref{th_sun_prof}. Examples of fitting several Gd, Dy and Th lines in the stellar spectra are presented in Fig. \ref{lines_prof}. 

\begin{table}
\caption{List of lines in the thorium 4019 \AA~line region }
\label{list_thline}
\begin{tabular}{lllcc}
\hline
Species & lambda, \AA &  Elow, ev & log gf & source \\
\hline
Ce II	&	4018.820	&	1.55	&	-0.959	&	VALD	\\
Nd II	&	4018.820	&	0.06	&	-0.890	&	VALD	\\
Fe I	&	4018.887	&	4.26	&	-2.781	&	solar fit	\\
Ce II	&	4018.900	&	1.01	&	-1.219	&	VALD	\\
Ce II	&	4018.927	&	0.63	&	-1.679	&	VALD	\\
V I	&	4018.929	&	2.58	&	-0.556	&	VALD	\\
Pr II	&	4018.963	&	0.20	&	-1.029	&	VALD	\\
13CH    & 4018.965      & 0.46     &   -3.253  &  VALD    \\
Mn I	&	4018.987	&	4.35	&	-1.883	&	VALD	\\
Fe I	&	4019.002	&	4.32	&	-2.700	&	solar fit	\\
Fe I	&	4019.042	&	2.61	&	-3.100	&	solar fit	\\
V II	&	4019.044	&	3.75	&	-1.231	&	VALD	\\
Ce II	&	4019.057	&	1.01	&	-0.529	&	VALD	\\
Mn I	&	4019.066	&	4.67	&	-0.522	&	VALD	\\
Ni I	&	4019.067	&	1.94	&	-3.399	&	VALD	\\
13CH    &   4019.074    &  0.46    &   -3.245  &  VALD    \\
Co I	&	4019.110	&	2.28	&	-3.287	&	VALD	\\
Co I	&	4019.118	&	2.28	&	-3.173	&	VALD	\\
Co I	&	4019.120	&	2.28	&	-2.876	&	VALD	\\
Co I	&	4019.125	&	2.28	&	-3.492	&	VALD	\\
Co I	&	4019.126	&	2.28	&	-3.298	&	VALD	\\
Th II	&	4019.129	&	0.00	&	-0.227	&	VALD	\\
Co I	&	4019.129	&	2.87	&	-5.163	&	VALD	\\
V I	&	4019.134	&	1.80	&	-2.149	&	VALD	\\
Co I	&	4019.135	&	2.28	&	-3.287	&	VALD	\\
Co I	&	4019.135	&	2.28	&	-3.474	&	VALD	\\
Co I	&	4019.138	&	2.28	&	-3.173	&	VALD	\\
Co I	&	4019.140	&	2.28	&	-3.298	&	VALD	\\
Co I	&	4019.143	&	2.87	&	-5.142	&	VALD	\\
Co I	&	4019.210	&	2.87	&	-4.821	&	VALD	\\
\hline  
\end{tabular}
\end{table}

In order to calculate the synthetic spectrum and the Th abundance, we used the relevant abundances of chemical elements obtained by \cite{mishenina:13}, including nickel. In particular, to take into account the blend due to cobalt, as a first approximation we estimated its abundance from the scaled solar cobalt value. Then, we further refined from the profile fit of the cobalt line at a wavelength of 4020.89 \AA, which was calculated factoring in the hyperfine structure (HFS). We finally derived the Th abundance by taking into account the contribution of cobalt in the blend Th-Co. Therefore, our results obtained for thorium should not be overestimated because of local contribution from other elements. Examples of fitting Co line in the stellar spectra are shown in Fig. \ref{th_prof}. 
The abundance of europium was determined by us early and for further analysis in this study, we use those obtained in \cite{mishenina:13}. In that study the Eu abundance was derived from the Eu II lines at 6645 \AA, taking into account the hyperfine structure \cite[][]{ivans:06}. 
The solar abundances of Dy, Gd and Th are determined using the STARSP code \cite[][]{tsymbal:96} from the lines in the spectra of the Moon and asteroids obtained with the ELODIE spectrograph with the line parameters being the same as in the stellar spectra: log A(Gd) = 1.08$\pm$0.05 and log A(Dy) = 1.10$\pm$0.05, which coincide with \cite{asplund:09} (log A(Gd)$_\odot$ = 1.07$\pm$0.04, log A(Dy)$_\odot$ = 1.10$\pm$0.04), and our solar log A(Th) = 0.08$\pm$0.08 is consistent with the value log A(Th)$_\odot$ = 0.08 reported for the Sun in \cite[][]{mashonkina:14b}, the value of  \cite{asplund:09} is log A(Th)$_\odot$ = 0.02$\pm$0.10. 

The stellar parameters and obtained Gd, Dy and Th abundances with the statistical uncertainties associated from line-to-line abundance variation (standard deviation or rms derivation) are given in Table \ref{ncapt}.

\subsection{Errors in abundance determinations}

The total errors in Gd, Dy, Th abundance determinations mainly result from the errors in sampling the parameter values and fitting the synthetic spectra to observational ones (0.05 dex in Gd and Dy, and 0.08 for Th). 
To determine the systematic errors in the elemental abundances, resulting from uncertainties in the atmospheric parameters, we derived the elemental abundance of 
four stars with different set of stellar parameters (\Teff~ in K, \logg, \Vt~ in km s$^{-1}$ , [Fe/H]): 
HD154345 (5503,4.30,1.3,-0.21), HD82106 (4827,4.10,1.1,-0.11), HD75732 (5373,4.30,1.1,0.25), and HD201891 (5850,4.40,1.0,-0.96)	
for several models with modified parameters 
($\Delta$\Teff = +100~K, $\Delta$\logg = +0.2, $\Delta$\Vt = +0.1). 
The impact of the parameter uncertainties on the accuracy of elemental abundance determinations, as exemplified by the stars with different \Teff~ and metallicities, is presented in Table \ref{errors}.

\begin{table}
\caption{
Abundance errors due to atmospheric parameter uncertainties, for four stars with different set of stellar parameters (\Teff, \logg, \Vt, [Fe/H]): 
HD154345 (5503,4.30,1.3,-0.21), HD82106 (4827,4.10,1.1,-0.11), HD75732 (5373,4.30,1.1,0.25), and HD201891 (5850,4.40,1.0,-0.96).	
}
\label{errors}
\begin{tabular}{lllccc}
\hline
& & HD1545345  && &  \\
 AN & El  & $\Delta$ \Teff+  & $\Delta$ \logg+ & $\Delta$ \Vt+ & tot+   \\                    
\hline
64	&GdII	&0.07 &0.10  &0.01	&0.12   \\ 
68	&DyII	&0.09 &0.08  &0.02	&0.13	  \\ 
90     &ThII&0.10& 0.11   & 0.0      &0.16   \\
& & HD82106  && &  \\
64	&GdII	&0.07 &0.06  &0.00	&0.10   \\ 
68	&DyII	&0.10 &0.12  &0.01	&0.15	  \\ 
90     &ThII&0.05& 0.06   & 0.0      &0.11   \\
& & HD75732  && &  \\
64	&GdII	&0.06 &0.10  &0.01	&0.12   \\ 
68	&DyII	&0.10 &0.09  &0.01	&0.14	  \\ 
90     &ThII&0.10& 0.06   & 0.0      &0.14   \\
& & HD201891  && &  \\
64	&GdII	&0.02 &0.04  &0.02	&0.08   \\ 
68	&DyII	&0.05 &0.03  &0.01	&0.07	  \\ 
90     &ThII&0.12& 0.06   & 0.0      &0.15   \\
\hline                             
\end{tabular}
\end{table}

As can be seen from Table \ref{errors}, uncertainties in \Teff~ and \logg~ contribute maximally to the total error.
Total errors due to parameter uncertainties and the measured spectra vary from 0.07 dex to 0.15 dex for Gd and Dy, and from 0.11 to 0.16 dex for Th abundance.

\begin{figure}
\begin{tabular}{ccc}
\includegraphics[width=\columnwidth]{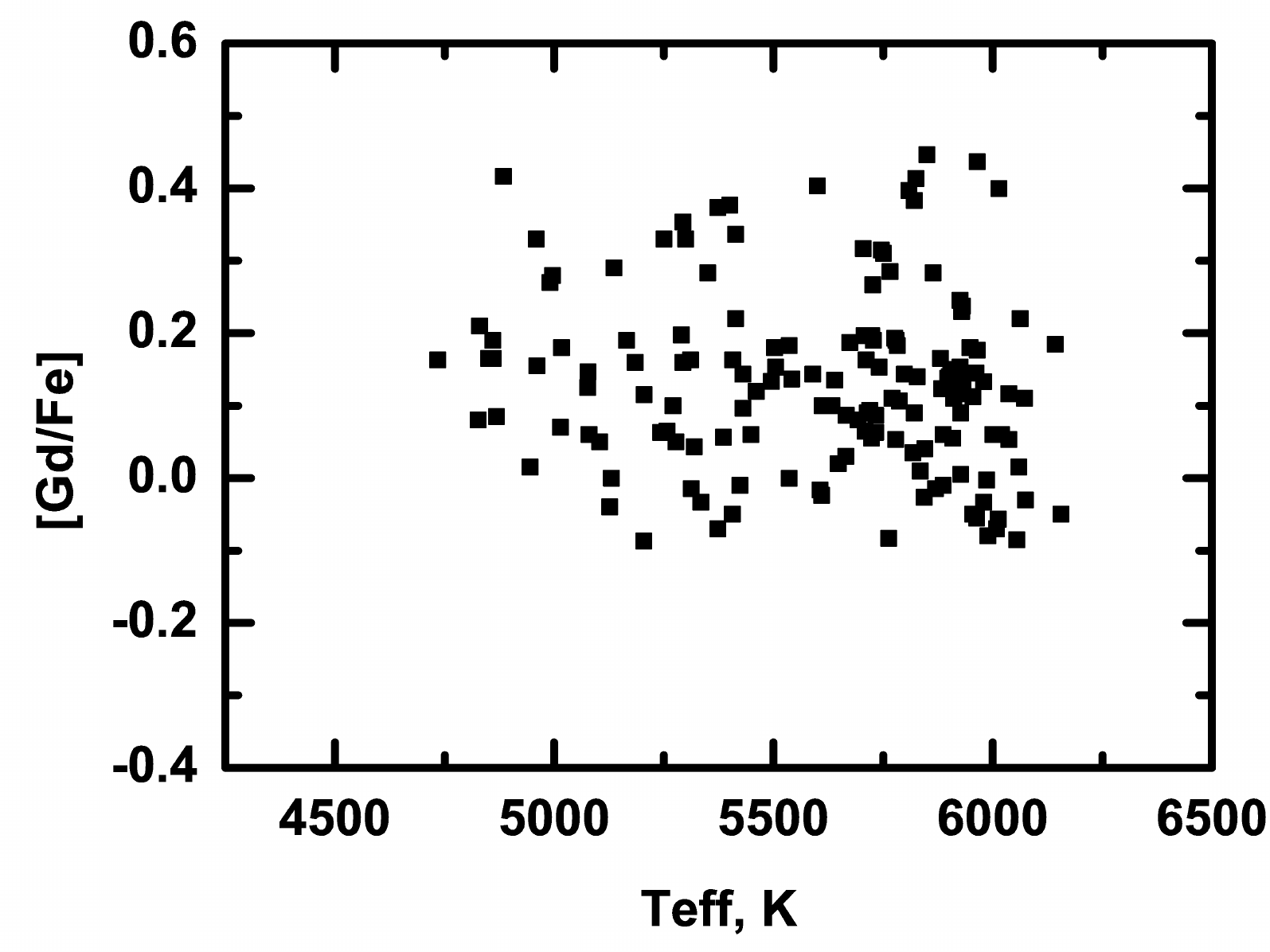}\\
\includegraphics[width=\columnwidth]{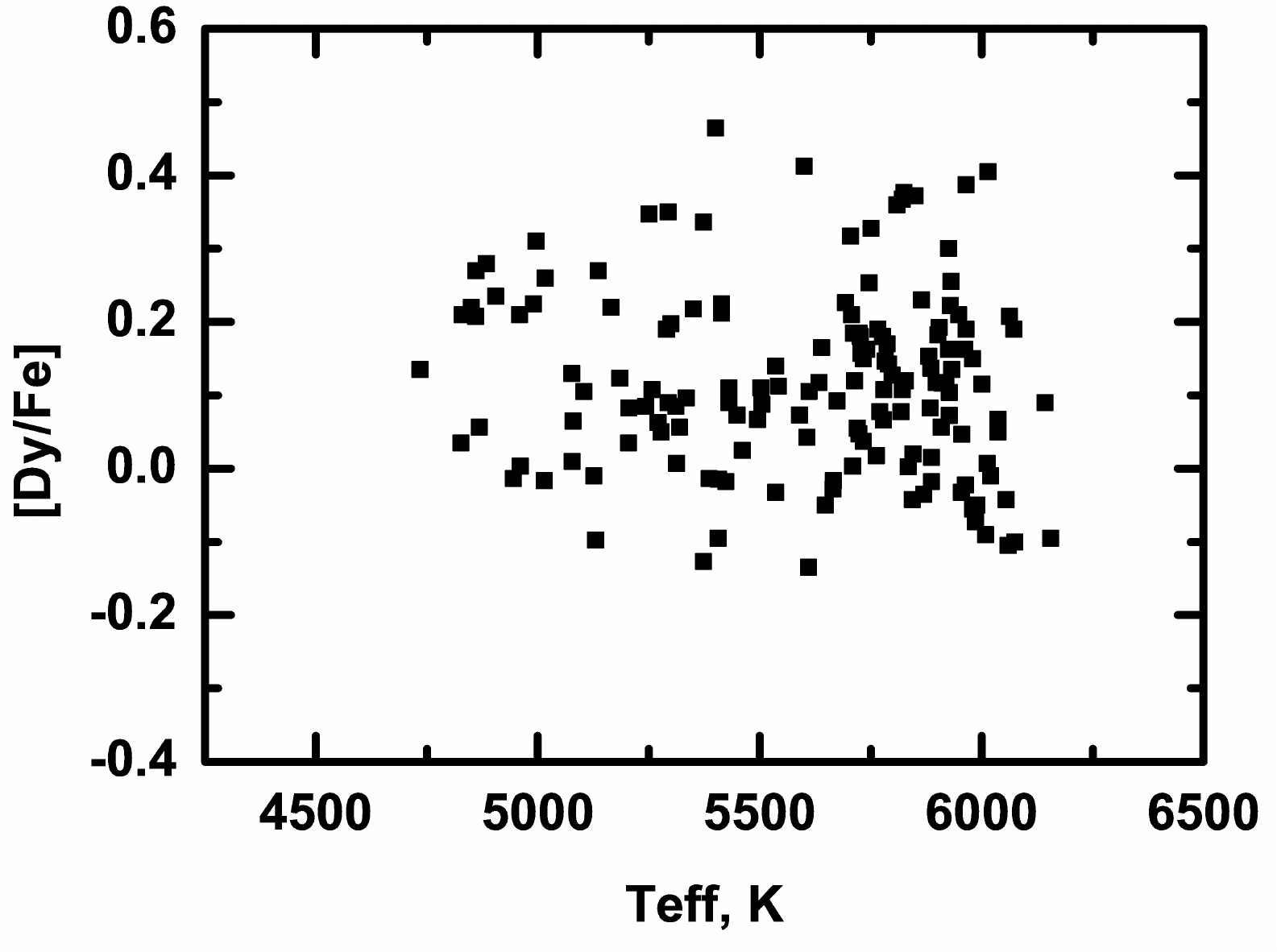}\\
\includegraphics[width=\columnwidth]{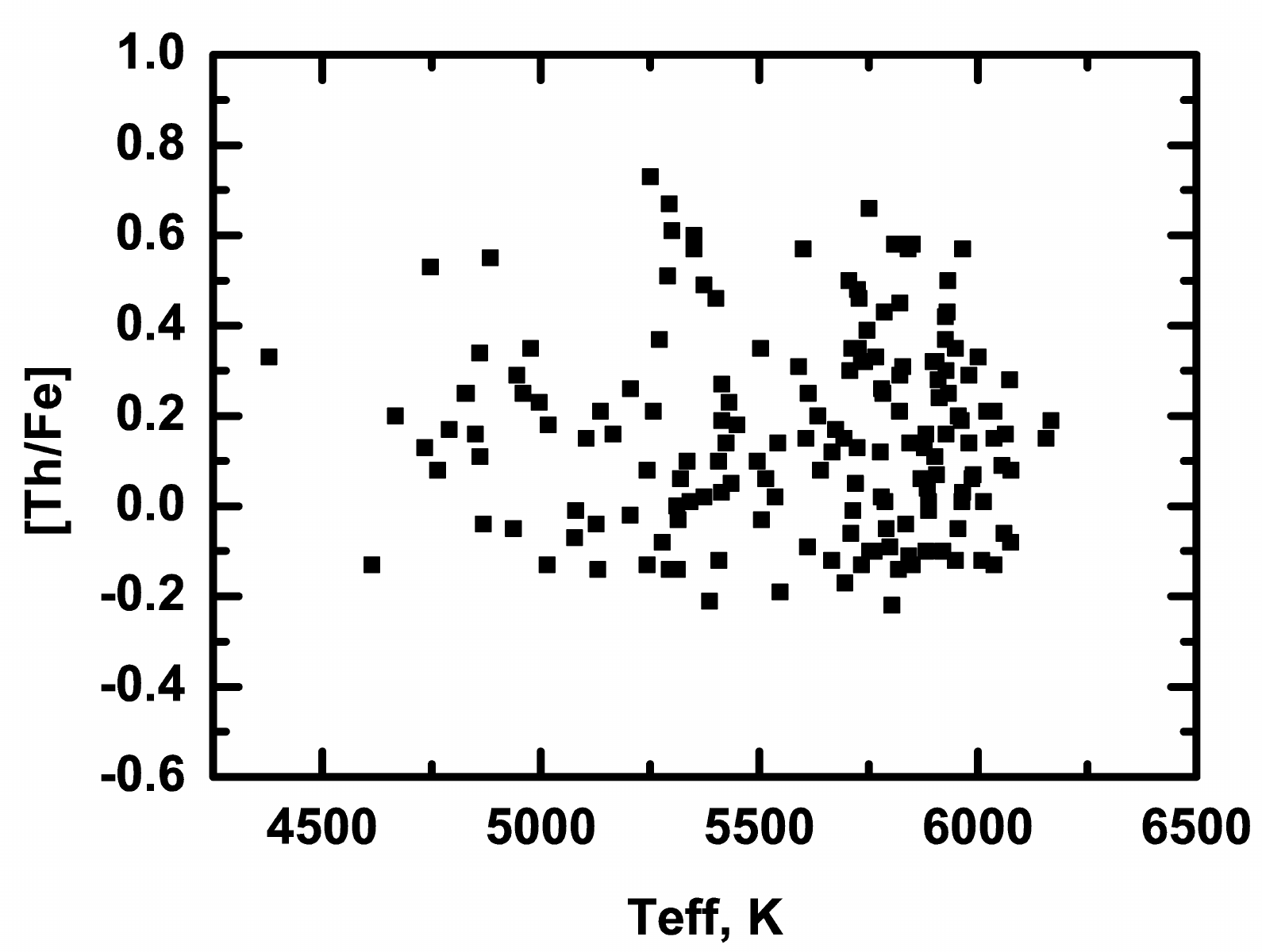}\\
\end{tabular}
\caption{Dependence of [El/Fe] vs. \Teff}
\label{el_teff}
\end{figure}

\begin{figure}
\begin{tabular}{ccc}
\includegraphics[width=\columnwidth]{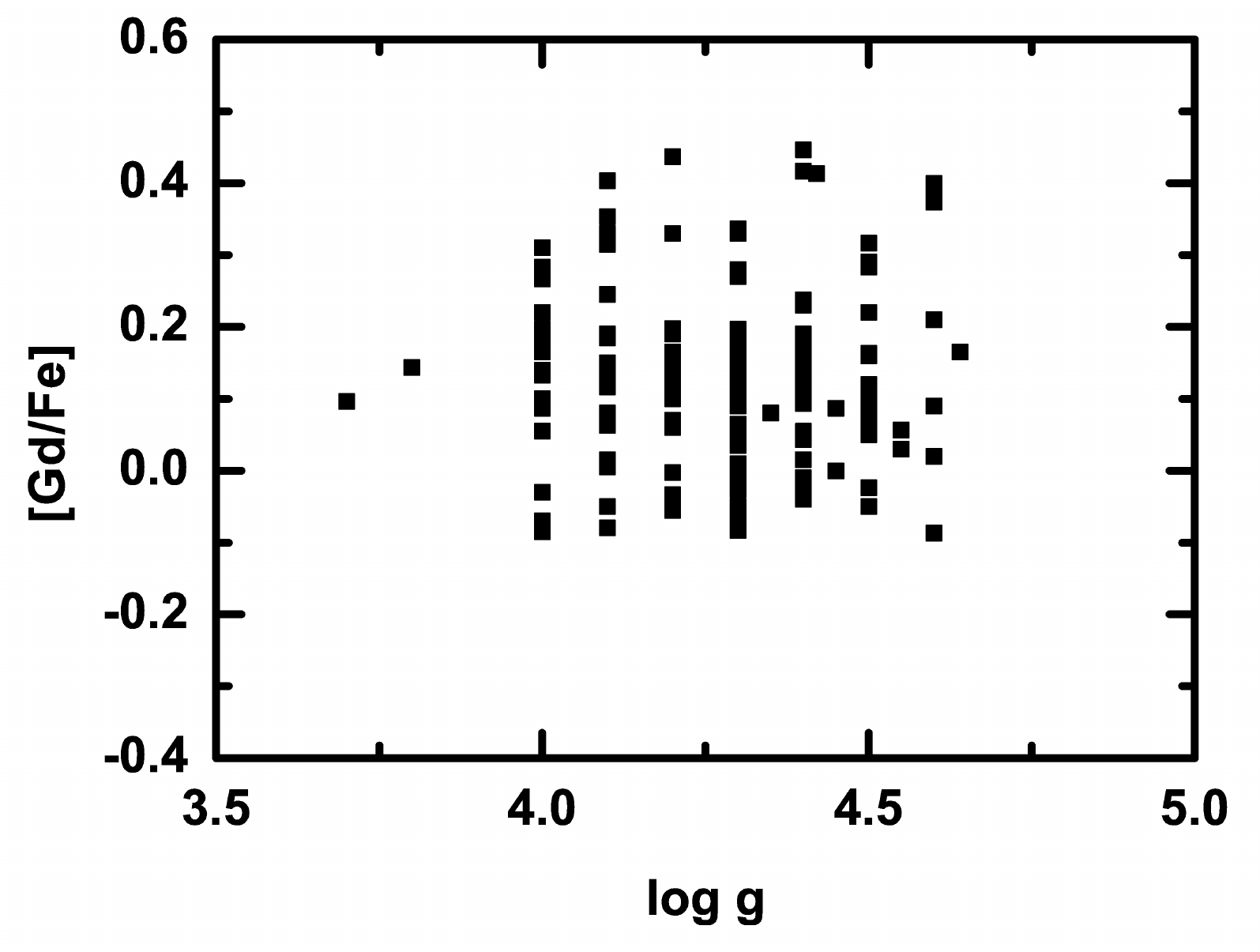}\\
\includegraphics[width=\columnwidth]{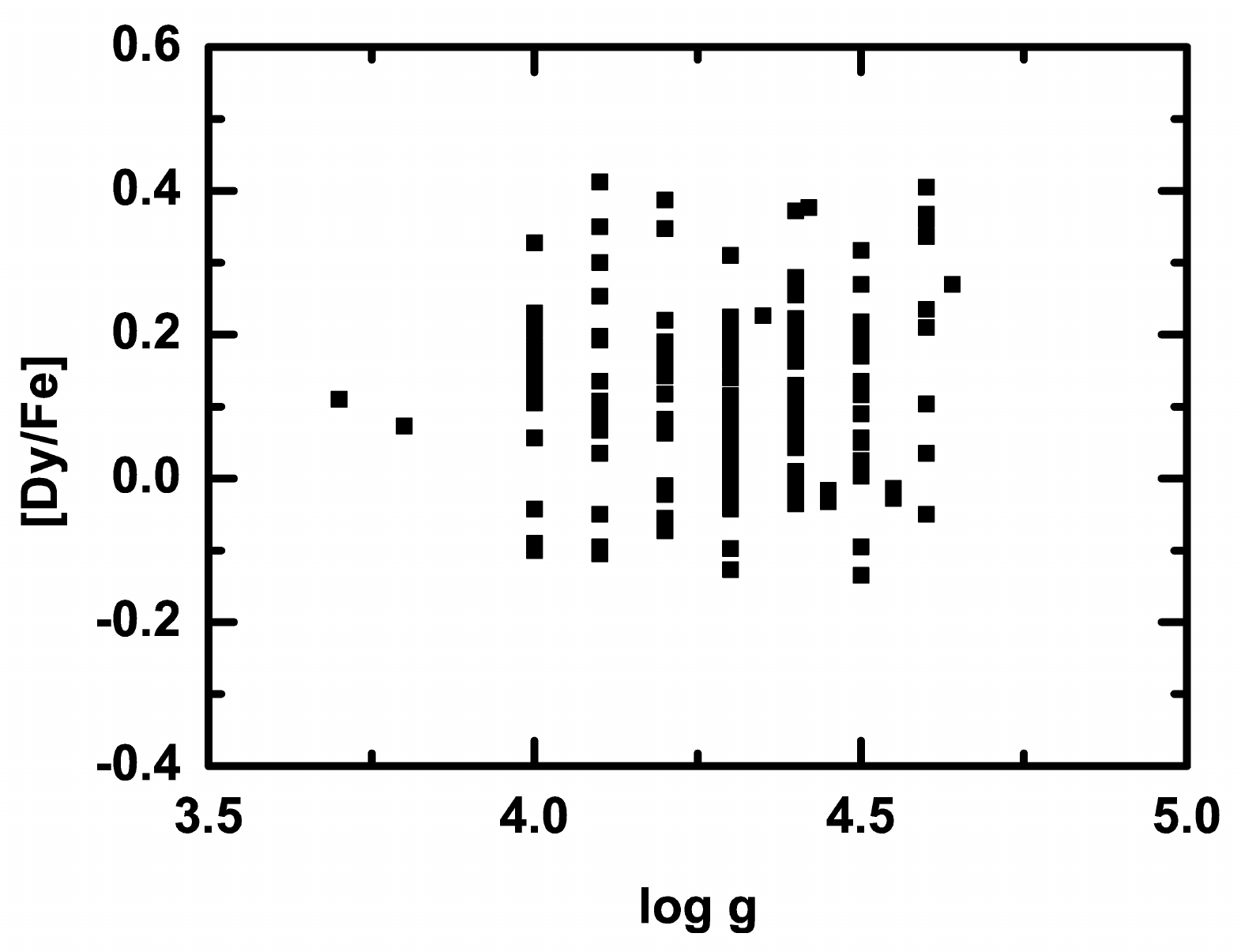}\\
\includegraphics[width=\columnwidth]{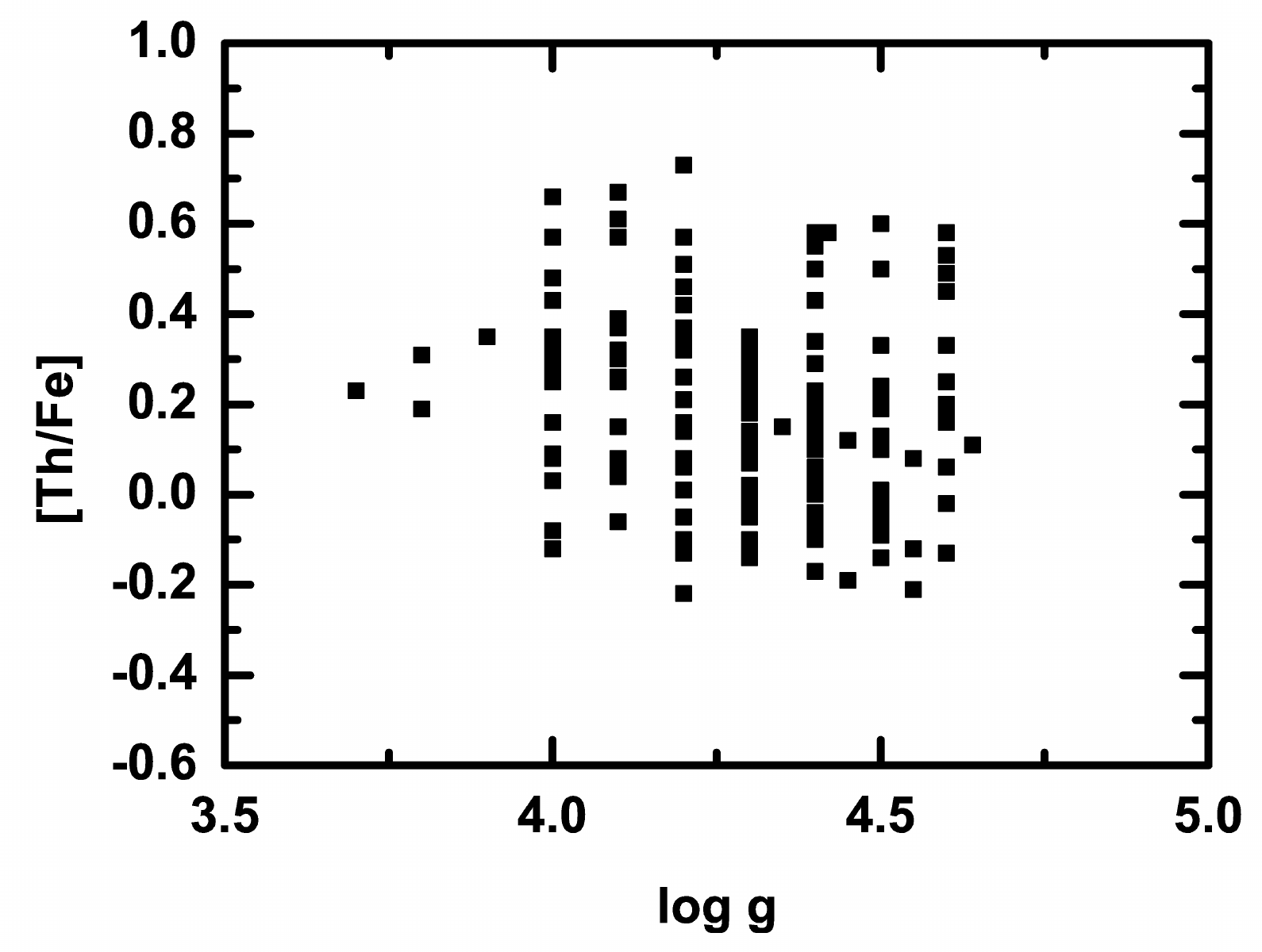}\\
\end{tabular}
\caption{Dependence of [El/Fe] vs. \logg}
\label{el_logg}
\end{figure}

To verify  our selection of stellar parameters, we present correlations between Gd, Dy and Th abundances and atmospheric parameters \Teff~ and \logg~(see Figs. \ref{el_teff}, \ref{el_logg}). As can be seen in Figures \ref{el_teff}, \ref{el_logg}, there is no correlation between the elemental abundances and chosen parameters. 

\begin{figure}
\begin{tabular}{cccc}
\includegraphics[width=\columnwidth]{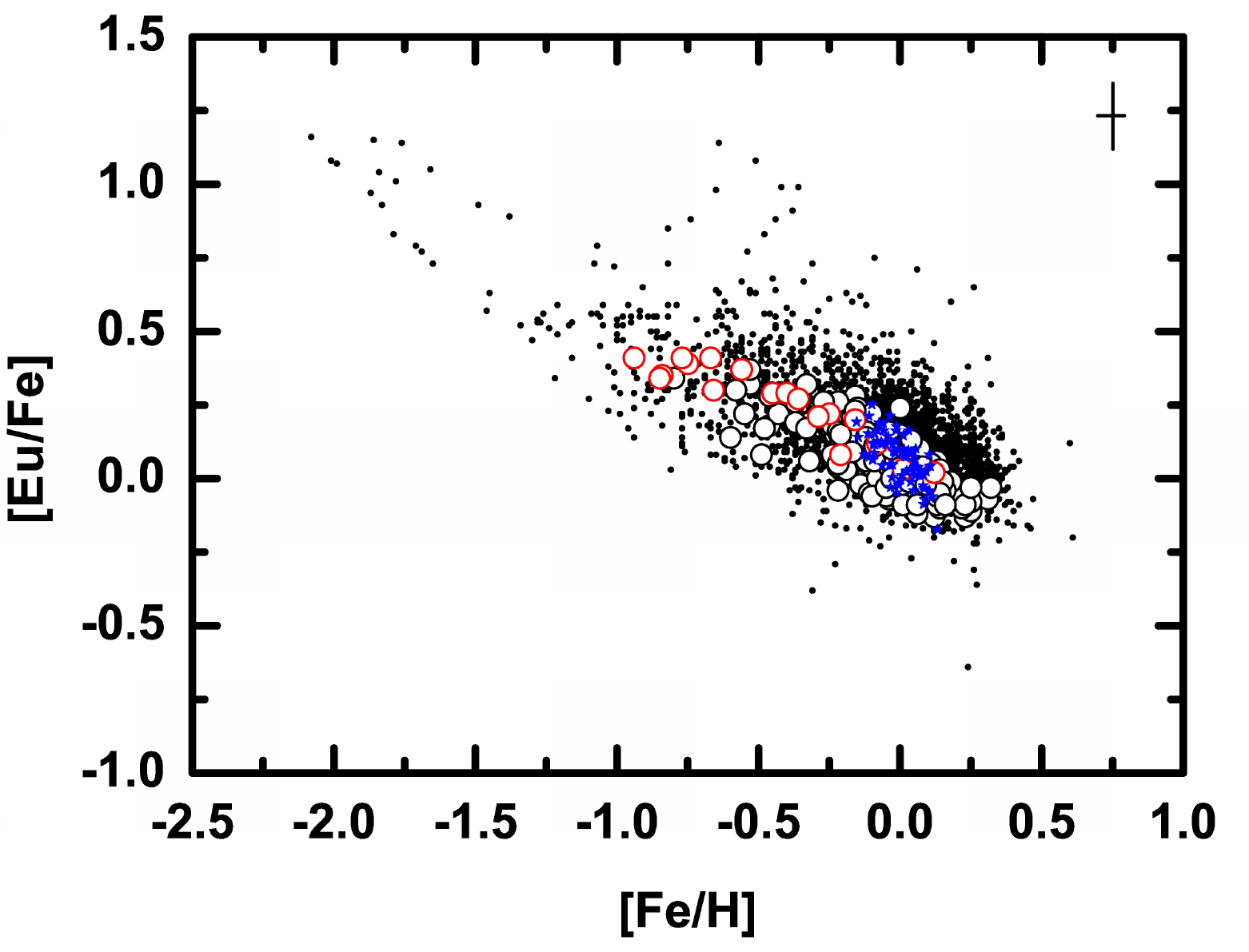}\\
\includegraphics[width=\columnwidth]{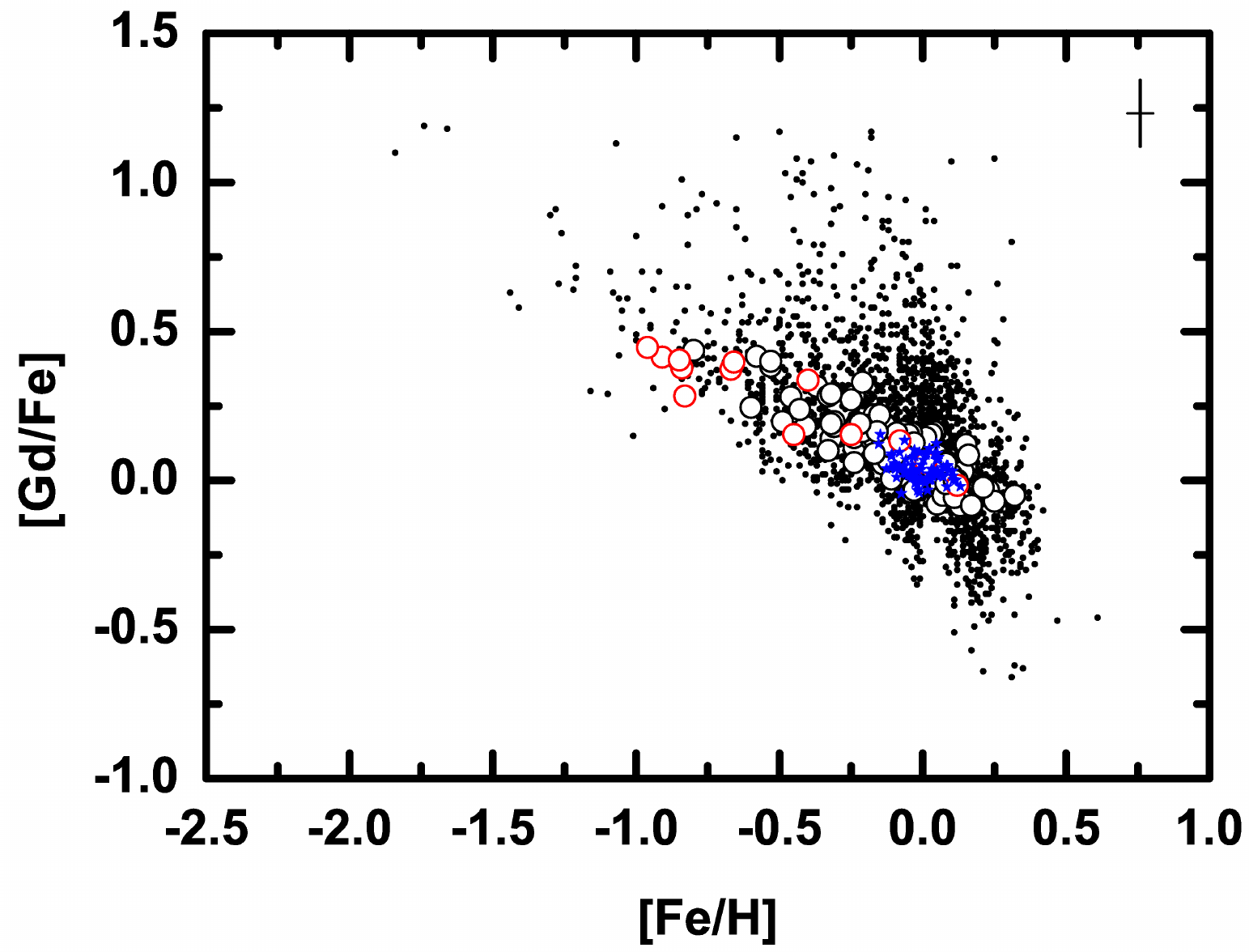}\\
\includegraphics[width=\columnwidth]{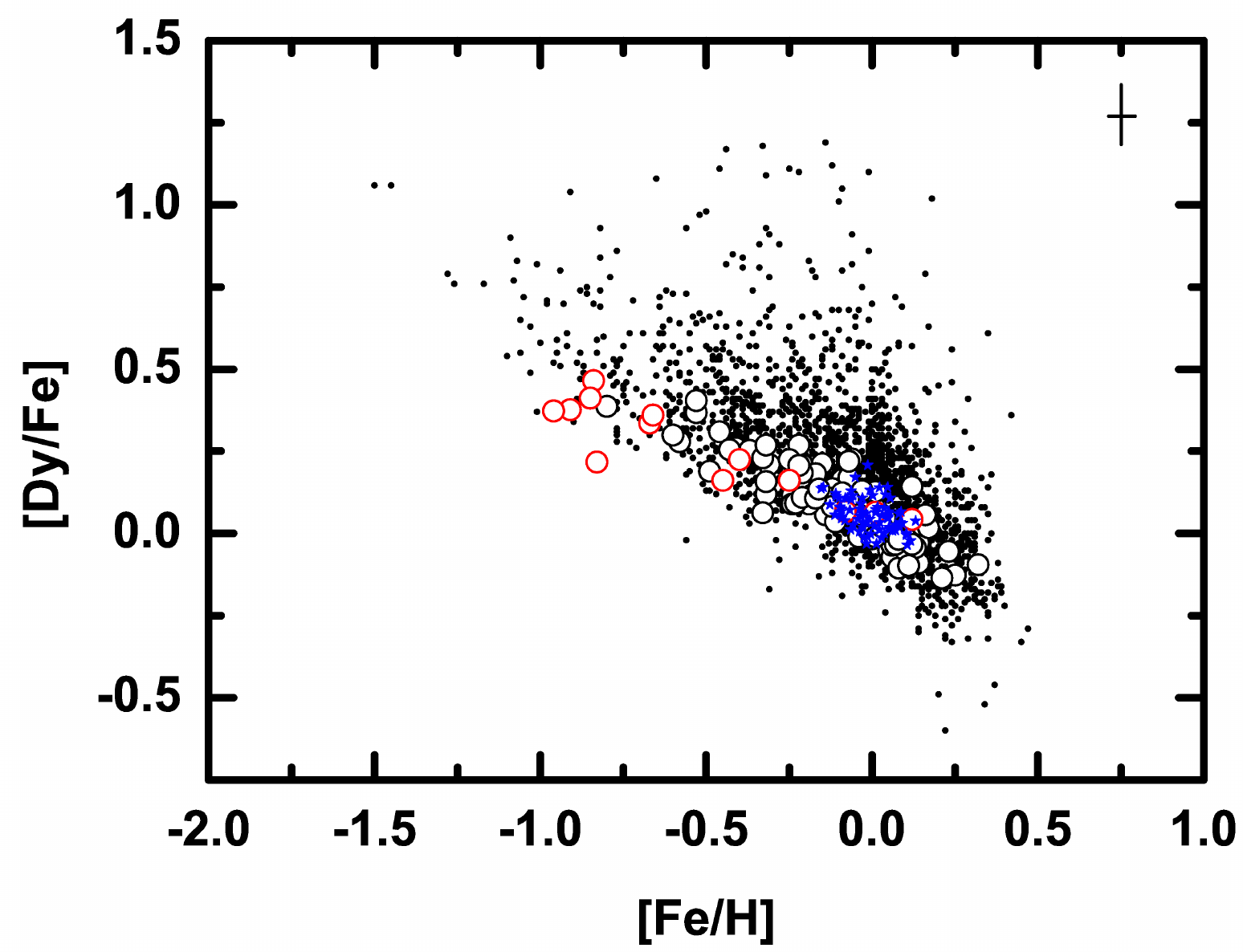}\\
\end{tabular}
\caption{[Eu/Fe], [Gd/Fe], and [Dy/Fe] as a function of [Fe/H]. Stars associated to the thin and thick discs are marked as black and red open circles, respectively. Selected data are taken from \protect\cite{guiglion:18} 
(points), and ones from \protect\cite{spina:18} 
(asterisks).}
\label{ba_eu_gd_dy}
\end{figure}

\begin{figure}
\begin{tabular}{c}
\includegraphics[width=\columnwidth]{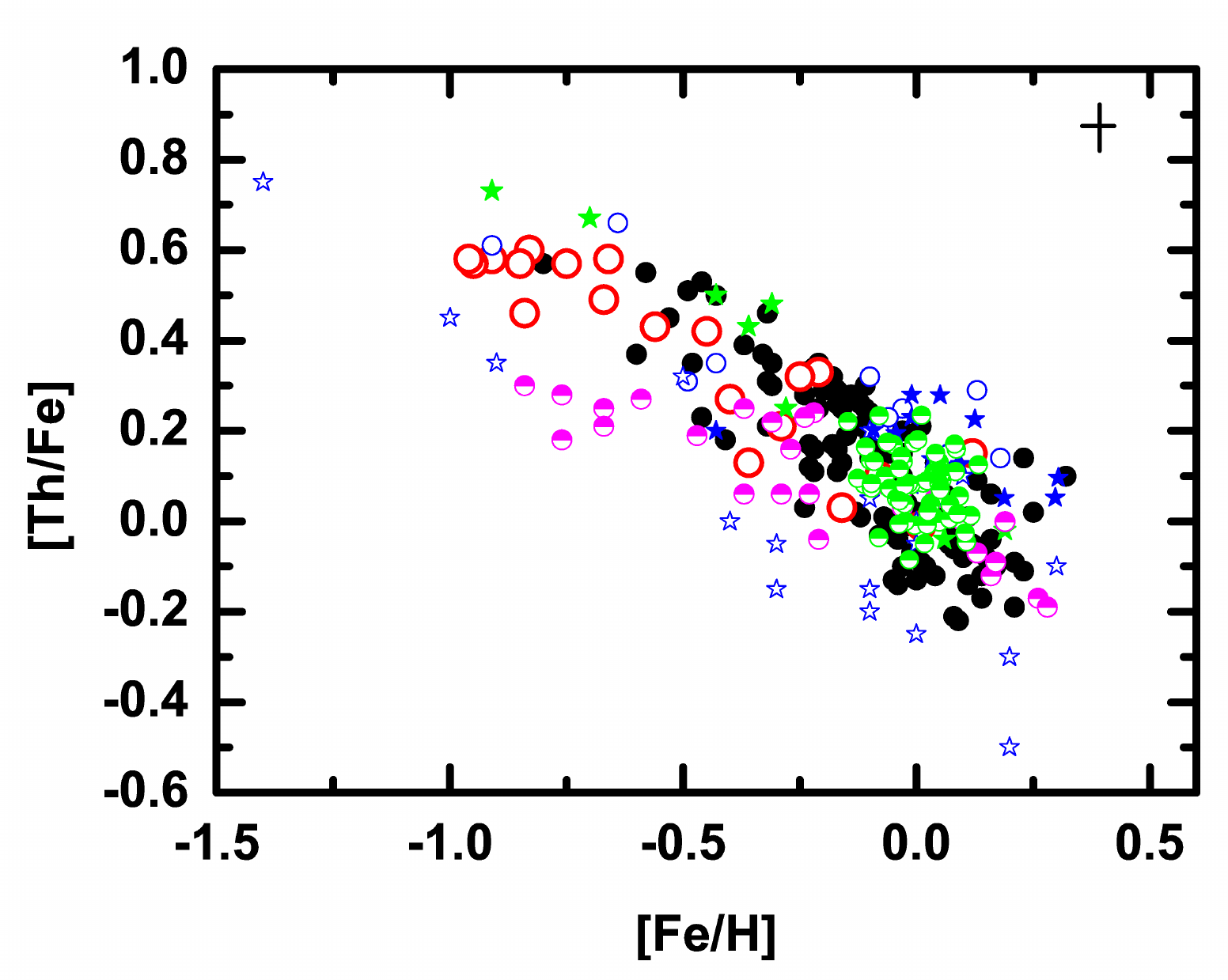}\\
\end{tabular}
\caption{[Th/Fe] as a function of [Fe/H]. Our measurements associated to the thin and thick discs are marked as black circles and red open circles, respectively. In comparison, observations by \protect\cite{peloso:05} (blue open circles), \protect\cite{morell:92} (blue open asterisks), \protect\cite{unterborn:15} (blue asterisks) and \protect\cite{botelho:19} (green semi-full circles) are shown.}
\label{comp_th}
\end{figure}

A comparison between the abundance determinations obtained in this study and the data reported by other authors is given in Table \ref{ncap} (see in  \S \ref{sec: stellar param}).
Also Fig. \ref{ba_eu_gd_dy} shows our [Eu/Fe], [Gd/Fe], and [Dy/Fe] data and ones from \cite{guiglion:18} and \cite{spina:18} as a function of [Fe/H]. 
For these Figures we selected the data of \cite{guiglion:18} with \Teff~ from 5100 K to 6300 K and surface gravities within 3.5 $<$ \logg $<$ 5.0. Such ranges of parameter values are the same chosen by the authors for further analysis.
The Fig. \ref{comp_th} presents our [Th/Fe] determinations and those by \cite{peloso:05}, \cite{morell:92}, \cite{unterborn:15} and \cite{botelho:19}. 

\begin{figure}
\begin{tabular}{cc}
\includegraphics[width=\columnwidth]{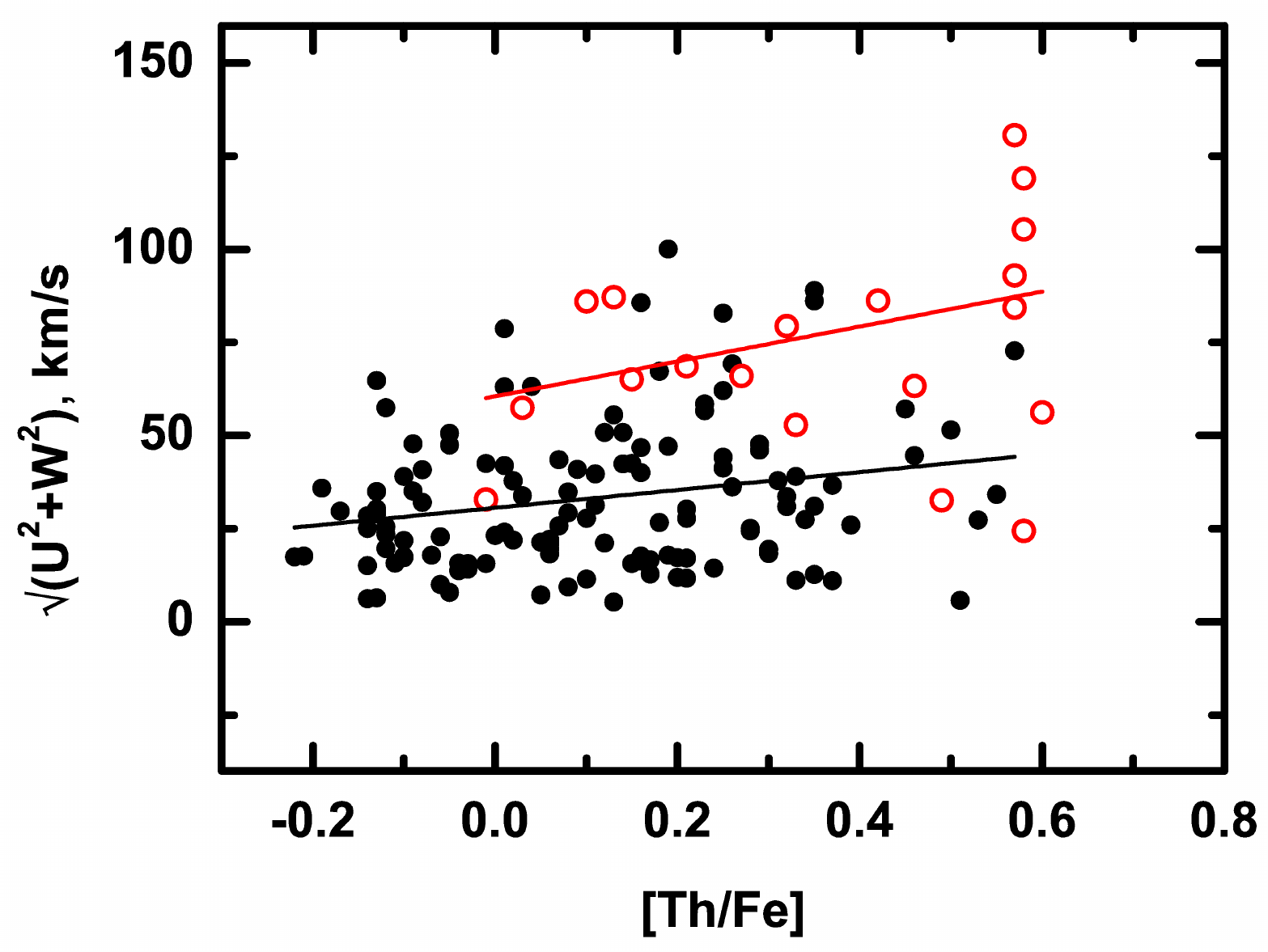}\\
\includegraphics[width=\columnwidth]{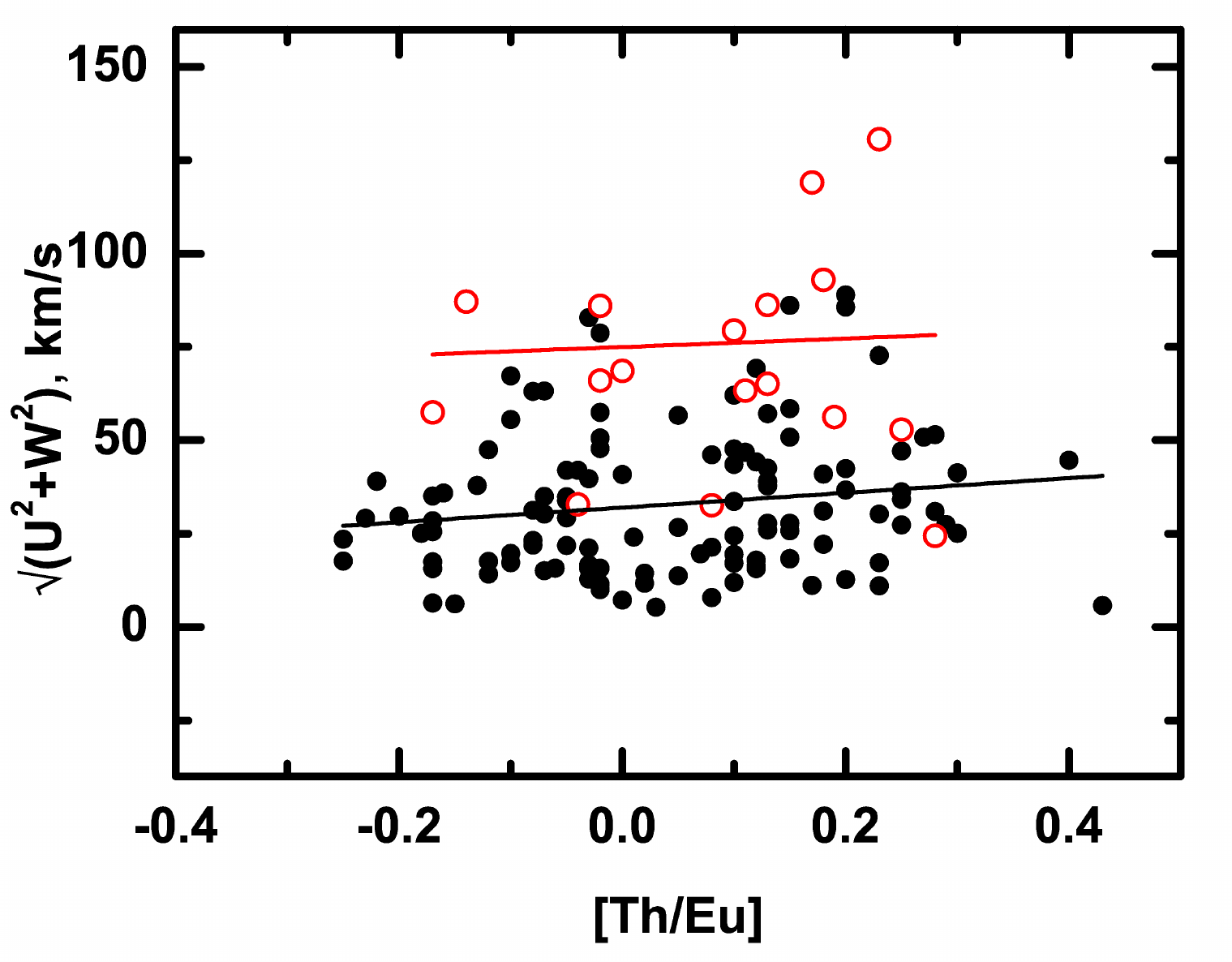}\\
\end{tabular}
\caption{For our stellar sample, the velocities along the Z axis ($\sqrt{U^2 + W^2}$) are shown with respect to [Th/Fe] and [Th/Eu]. The full sample is divided between thick and thin disk stars. 
}
\label{fig: thvsage_thvsvelocity}
\end{figure}

\begin{figure}
\begin{tabular}{c}
\includegraphics[width=\columnwidth]{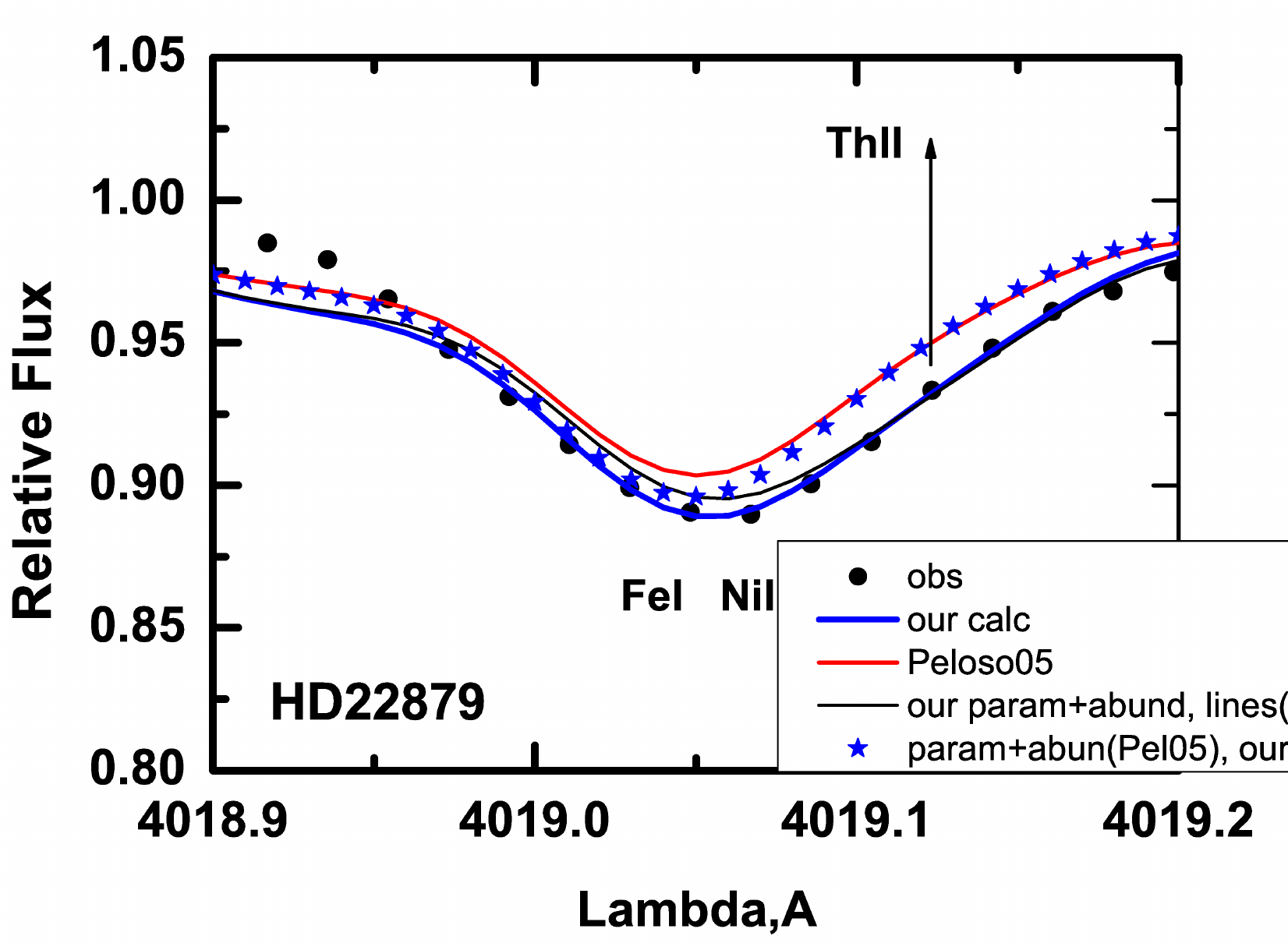}\\
\includegraphics[width=\columnwidth]{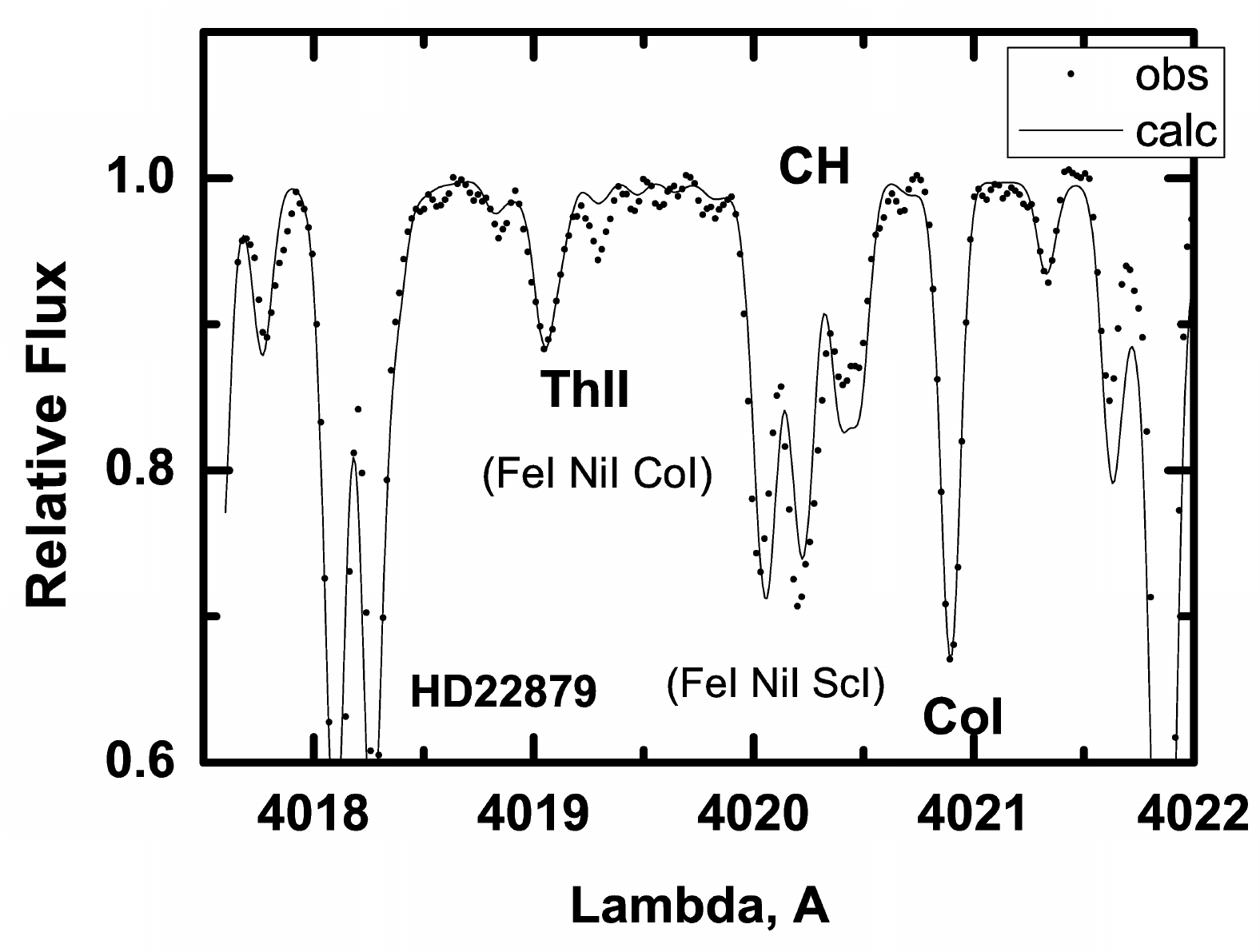}\\
\end{tabular}
\caption{Top panel: comparison of the synthetic spectrum calculations for our data and those of Peloso et al. (2005)(see details in the text); bottom panel: comparison of our synthetic computation  with  observations, in the region of Th II line for a wider range of the spectrum. }
\label{th_prof}
\end{figure}

Table \ref{ncap} shows the mean differences and rms errors for the values of thorium abundance obtained by us and other authors for stars in common. We have 5 common stars with \cite{morell:92}, for which the parameters and content of thorium are determined: $\Delta$([Th/Fe]) (our – Morell) = 0.11 $\pm$0.12 dex, the shift is within the error limits.  We also have 5 stars in common with \cite{peloso:05}, for which parameters have been determined, and 4 of them have definitions of thorium abundance. The mean difference is $\Delta$([Th/Fe]) (our-Peloso)= 0.23 $\pm$0.14 dex and it is larger than the resulting error.  In addition, there is one star HD 76932 also shared with \cite{morell:92} and \cite{peloso:05}. Our thorium abundance for HD 76932 is [Th/Fe] = 0.57 dex,   \cite{morell:92} and \cite{peloso:05} give [Th/Fe] = 0.35dex  and [Th/Fe] = 0.30 dex, respectively.  
There is  one common star, namely HD 146233, with  \cite{unterborn:15}, it is also was studied by \cite{botelho:19}. For HD 146233 the Th measurements are varying significantly between different works. We obtain [Th/Fe] =--0.09 dex,   \cite{unterborn:15} gives [Th/Fe] =0.28 dex and in \cite{botelho:19} [Th/Fe] =0.15 dex.
We have 5  stars in common  with the \cite{botelho:19}, but among them only 3 stars with measurements of Th abundance. The mean difference and rms  between our data and those of \cite{botelho:19} is $\Delta$([Th/Fe]) (our – Botelho) = -0.09 $\pm$0.15. This value (shift) has opposite sign in comparison with those obtained for comparison  of \cite{morell:92}, and \cite{peloso:05}, but it is within the limits of errors. 
From a comparison with \cite{morell:92}, \cite{peloso:05}, and \cite{botelho:19}, the obtained rms of the mean difference are 0.12, 0.14 and 0.15, respectively.  
At the same time, the \cite{peloso:05} data show a systematic shift relative to our data (0.23), since the mean  difference is greater than the scatter. The maximum difference in thorium abundance between our values and these for \cite{peloso:05} reaches 0.4 dex for the star HD 22879. However, this discrepancy is contributed also by a difference of 0.15 dex in metallicity obtained by us ([Fe/H] = -0.91) and ([Fe/H] = -0.76)  \cite[][]{peloso:05}. 
Fig. \ref{th_prof} (top panel) presents the synthetic spectra for the star HD 22879 calculated in this work (blue solid line) and that with the data (stellar parameters, chemical abundances and line list) of \cite[][]{peloso:05} (red solid  line).
 The black solid line shows the calculation from our data (parameters, abundance), but adapted to the  Peloso's line list. Blue asterisks shows the calculation based on Peloso's data (parameters, abundances) with our line list in the thorium region. Circles are the corresponding observational spectrum.
 The bottom panel shows a description of the observed spectrum by our synthetic spectrum in a wider spectral region with the thorium line. 
We observe a difference in the synthetic calculations for our data and those of \cite[][]{peloso:05} in the region of the iron, nickel, manganese lines at the maximum intensity  of this spectral peculiarity due to the difference in oscillator strengths of lines, assumed abundances of elements and the absence of the iron line in the \cite[][]{peloso:05} list. In the part of profile with the thorium-cobalt lines, we see that using our data and different line lists gives similar trends, and different stellar parameters and abundances make a significant contribution to the result, in this case, an increase in the thorium abundance compared to that obtained by \cite[][]{peloso:05} is required.
In general, the differences between the data obtained in different works are mostly due to the different lists of used lines in the thorium line region and different parameters and elemental abundances measured in different works.

\section{Results and comparison with Galactic chemical evolution models} 
\label{sec: result, gce}

The Milky Way disk stars with the metallicity range covered from this study are formed from interstellar matter enriched by several generations of stars. Therefore, these observations cannot be directly compared with theoretical stellar models, and Galactic chemical evolution (GCE) simulations must be used to study the evolution history that allowed to build the chemical inventory observed today \citep[e.g.,][and references therein]{tinsley80,gibson03,kobayashi:20,matteucci:21}. In this work we focus on the r-process elements Eu, Gd, Dy and Th. Although the main stellar source of the r-process in the Galaxy is still matter of debate, from decades the r-process was typically considered to produce the same abundances independently from the metallicity of the stellar progenitors. The close similarity between the solar residual \citep[where the residual is derived by subtracting the s-process contribution from the solar abundances of heavy elements beyond iron, e.g., ][]{arlandini:99,bisterzo:14,prantzos20} and the abundance patterns measured in r-process-rich metal-poor stars drove and supported such a scenario. Indeed, while there is a significant abundance scatter among different r-process enriched stars for lighter elements in the mass region Sr-Ru, the abundances appear to better align with the solar residual for Ba and heavier elements up to Pb \citep[e.g.,][]{sneden:08, cowan:21}. 
Within this heavier mass region, GCE simulations could carry the same r-process signature across the evolution of the Galaxy, where the main remaining uncertainties are the source frequency and the quantitative r-process yields associated to each stellar source (e.g., \citealt{matteucci:14,wehmeyer:15,coteLIGO,hotokezaka:18}).
A larger observational scatter between different r-process rich stars has been measured for the actinides elements Th and U, which are highly affected by varying the conditions in r-process theoretical calculations \citep[][]{eichler:19, cowan:21}. 
However, a significant variation is also seen beyond Ba once a larger sample of metal-poor stars is considered \citep[][]{roederer:10}, suggesting that the r-process production does not yield a unique and robust pattern, and a degree of variation should be expected. The observation of actinide-boost stars has further questioned those classical paradigms \citep[e.g.,][]{roederer:10, holmbeck:18,farouqi:21}.

At least for metal-poor halo stars in the Galaxy, it is still matter of debate if only one r-process source would be able to explain the early large variations observed in stars for Eu and other heavy r-process element abundances with respect to iron or $\alpha$-elements. As already pointed out by \citet{Qian.Wasserburg:2007} and followed up by \citet{Hansen.Montes.Arcones:2014}, it is actually more plausible that at least two different types of r-process sources were active, contributing with different frequency and timescale. Here we discuss the aspect that two processes contribute to the lanthanide and actinide r-process elements. On the other hand, observations from old metal-poor stars would not exclude that today there is one source with possible abundance variations, dominating the r-process contribution to GCE \citep[e.g.,][and references therein]{wehmeyer:15,cote:19,farouqi:21}.

If we consider the r-process lanthanides discussed in this work, i.e., Eu, Gd and Dy, we have seen in Figure~\ref{ba_eu_gd_dy} that their abundance trends with respect to Fe are similar. In particular, we cannot identify if the observed abundance scatter is due to the GCE contribution from multiple r-process sources and/or some different production in the region, or if such a dispersion can be simply due to observation uncertainties. 

On the other hand, it may be interesting to study the evolution of Th (an actinides element) with respect to Eu. In this context, we have performed GCE models to compare with our new observations. The simulations are made using the Python code \texttt{OMEGA+} (\citealt{coteomega,coteomegap}), which is part of the open-source JINAPyCEE package\footnote{https://github.com/becot85/JINAPyCEE}.  
It consists of a two-zone model that includes a one-zone GCE model surrounded by a large gas reservoir representing the circumgalactic medium. These two zones are interacting via galactic inflows and outflows, where inflows transfer gas from the circumgalactic medium to the central GCE model (the galaxy), and outflows transfer gas from the galaxy to the circumgalactic medium. In this work, we use the yields of \cite{nomoto13}, \cite{cristallo:15}, and \cite{iwamoto99} for massive stars, low-and intermediate-mass stars, and Type~Ia supernovae, and we use the same galaxy evolution parameters as the best model found in \cite{cote19radio}, which reproduced various observational constraints such as the current star formation rate, gas inflow rate, supernova rates, total stellar mass, and total gas mass. Recent developments have allowed to take into account radioactivity throughout the GCE calculations (\citealt{cote19radio,2022ApJ...924...10T}), using the numerical solver presented in \cite{yague22} to properly follow radioactive decay on timescales shorter than the lifetime of the Milky Way. This numerical solver was modified in OMEGA+ to include the terms for material moving between the two simulated zones along with decay in an unsplit fashion.

\begin{figure*}
\begin{tabular}{c}
\includegraphics[width=\textwidth]{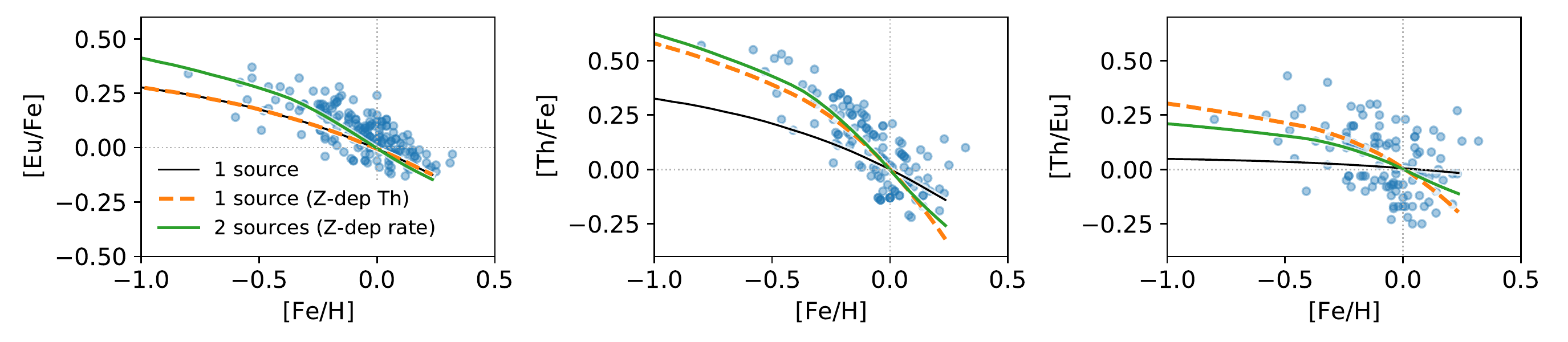}\\
\end{tabular}
\caption{Predicted evolution of [Eu/Fe], [Th/Fe], and [Th/Eu] as a function of [Fe/H] in the Galactic disc, using \texttt{OMEGA+}. In each panel, the solid black line assumes one r-process site with a prompt delay, assuming constant Eu and Th yields for all enrichment events. The dashed orange line is the same as the black line, but assuming Z-dependent Th yields. The solid green line assumes two r-process sites, one with a prompt delay and one with a long-lasting delay-time distribution, where the contribution of the prompt site declines with increasing metallicity.}
\label{fig:gce_v2}
\end{figure*}

Our results are shown in Figure~\ref{fig:gce_v2}, and our goal is to address the relative timescales at which Eu (as a representative of the r-process production of lanthanides including Dy and Gd) and Th (as a representative of the actinides) are produced within the Galactic disc with a simple approach. Given this goal, all Eu and Th yields in our models have been included artificially in order to freely explore which scenarios could give rise to the GCE trends provided by our data. The black solid lines assume that all Eu and Th come from one source, with a short delay time typical of CCSNe sources. Here Eu and Th yields are the same in all events, regardless of metallicity. In this case, while our prediction is acceptable for [Eu/Fe], the predicted trend for [Th/Fe] does not decrease as steeply as the observational data, a feature that can also be seen with [Th/Eu]. The relatively flat [Th/Eu] trend shows that the decay of Th (the $^{232}$Th half-life is 14.1 billion years) plays an insignificant role in shaping our predictions, meaning that Th decay would not explain the different degrees at which Th and Eu decrease with respect to metallicity.

The dashed orange and solid green lines in Figure \ref{fig:gce_v2} explore two different scenarios matching the Th and Eu trends simultaneously. The dashed orange line still assumes one prompt r-process source and a constant yield for Eu, but assumes metallicity-dependent Th yields where Th is boosted by a factor of 4.5 at low metallicity relative to high metallicity, with a continuous decrease between $Z=0.001$ and 0.02. The solid green line, on the other hand, combines two r-process sources — neutron star mergers with long delay times, and exotic SNe or collapsars with short delays (see also e.g., \citealt{cote:19,haynes:19,siegel:19,molero21, farouqi:21}). In this case, Th and Eu yields are kept constant as a function of metallicity for both sources. However, the frequency of the short-delay source is assumed to be metallicity dependent, such that its rate is three times higher than the long-delay source at $Z<0.001$, and becomes negligible $Z>0.01$. Such a computational experiment would boost the Th production at low metallicity as we may expect. A type of exotic SNe that would fit these assumptions are magneto-rotational (MHD) supernovae. MHD supernovae were originally proposed as a source of a strong r-process \citep[e.g., ][]{winteler:12}. However, it was later shown by \cite{moesta:18} that a strong r-process can only be obtained with (unlikely) very extreme pre-collapse magnetic fields, which are required to eject neutron-rich matter stemming from the electron capture during the collapse to high densities. If sufficient rotation exists also weaker pre-collape magnetic fields can be enhanced by the magneto-rotational instability (MRI), lead to a successful explosion and a highly magnetized neutron star (magnetar), but during the delay encountered before the MRI had its impact,  neutrino absorptions enhance the electron fraction and limit the reach of the r-process. On the other hand, black-hole accretion disk outflows can lead to highly r-process enriched matter with $Y_e$-values in the ejecta just in the range leading to an actinide boost, as observed in many r-II stars \citep[see the detailed discussion in][]{farouqi:21}. As the collapsar behavior leading to black holes requires the core-collapse of quite massive progenitors, their frequency is expected to be much higher at low metallicities. The reason is that low-metallicity  progenitors have lower opacities, experience - as a consequence - less mass loss during their stellar evolution and possess at the point of core collapse significantly higher masses, favoring the collapse to a black hole.

Both scenarios shown in Figure~\ref{fig:gce_v2} lead to an enhancement of Th production in the early Universe, and would both be consistent with the observation of metal-poor actinide-boost stars. From this timescale experiment, it is unfortunately not possible to distinguish between one r-process source with metallicity-dependent yields, multiple r-process sources with metallicity-dependent rates, or a combination of the two. The variations observed in metal-poor stars between r-process elements and the existence of the actinide-boost stars seem to point more toward the second or third scenarios mentioned above \citep[e.g.,][]{roederer:10,farouqi:21} that, as we have seen,  it would be consistent with observations at higher metallicities in the Galactic disk.    
Nevertheless, the fact that [Th/Fe] decreases more steeply than [Eu/Fe] suggests that Th and Eu had a different production history, with Th being more efficiently synthesized at low metallicity than at high metallicity, as compared to Eu.

\section{Conclusions}
\label{sec: conclusions}

In this work, we presented and discussed the abundance measurements of Gd, Dy and Th for 276 disc stars. The analysis is based on LTE assumptions. 
Typical uncertainties are 0.10 dex for Gd (with a range between 0.08 and 0.12 dex), 0.11  for Dy (between 0.07 and 0.15 dex) and 0.12 dex for Th (with a range between 0.09 and 0.15 dex). The major sources of these uncertainties in the analysis are the stellar surface temperature and gravities. 

The [Dy/Fe] and [Gd/Fe] ratios show the same trend of [Eu/Fe] in the galactic disk. Due to the present observation uncertainties, it is not possible to use the evolution of Dy and Gd with respect to Eu to disentangle a contribution from different r-process components. On the other hand, [Th/Fe] shows a steeper decrease than the [Eu/Fe] with respect to [Fe/H].

By using GCE models we have explored possible solutions to explain those trends. We found that the observations may be better reproduced by one r-process source but with metallicity-dependent Th yields, or by multiple r-process sources with metallicity-dependent rates for the Th-rich source. We would rather support this second scenario, since it would also be compatible with observations of both actinides boost r-process rich metal-poor stars and not boosted.

\section*{Acknowledgements}

This  article  was  based  on  observations  collected  at  OHP  Observatory, France.
MP acknowledges significant support to NuGrid from NSF grant PHY-1430152 (JINA Center for the Evolution of the Elements) and STFC (through the University of Hull's Consolidated Grant ST/R000840/1), and access to {\sc viper}, the University of Hull High Performance Computing Facility.
MP acknowledges the support from the "Lendület-2014" Programme of the Hungarian Academy of Sciences (Hungary).
FKT acknowledges support from the European Research Council (FP7) under ERC Advanced Grant Agreement 321263 FISH.
BC, AY and MP acknowledges support from the ERC Consolidator Grant (Hungary) funding scheme (project RADIOSTAR, G.A. n. 724560) and from the National Science Foundation (USA) under grant No. PHY-1430152 (JINA Center for the Evolution of the Elements). This article is based upon work from the ChETEC COST Action (CA16117), supported by COST (European Cooperation in Science and Technology).
We thank the ChETEC-INFRA project funded from the European Union’s Horizon 2020 research and innovation programme (grant agreement No 101008324), and the IReNA network supported by NSF AccelNet. MP also thank the UK network BRIDGCE. TM is grateful to the Laboratoire d’Astrophysique de l’Universite de Bordeaux for their kind hospitality. 

\section*{Data Availability} 

The data underlying this article will be shared on reasonable request to the corresponding author.

\bibliography{proc}

\appendix

\section{}
We presented the stellar parameters and the Gd, Dy, Th  abundances with errors  in
Table A1.

\onecolumn
\clearpage

\begin{longtable}{lcccccccccc}
\caption{Stellar parameters and abundances of Gd, Dy, and Th.}
\label{ncapt}\\
\hline
HD/BD    & \Teff, K & \logg  & [Fe/H] & \Vt &  [Gd/Fe]& stand. deviation & [Dy/Fe] & stand. deviation & [Th/Fe]\\ 
\hline  
thin disc&  &  &  &  &  &  &  &    \\
\hline                               
\endfirsthead
\hline
HD/BD    & \Teff, K & \logg  & [Fe/H] & \Vt &  [Gd/Fe]& stand. deviation & [Dy/Fe] & stand. deviation   & [Th/Fe]\\ 
\hline  
\endhead
\hline                               
 166   &  5514 &   4.6  &   0.16 &   0.6  &--	&--     & --	& --    & 0.06  \\
 1562  &  5828 &    4.0   &  -0.32 &   1.2  &0.14	& 0.04	& 0.12	& 0.05	& 0.31  \\
 1835  &  5790 &   4.5  &   0.13 &   1.1  &--	& --	& --	& --	& --      \\
 3651  &  5277 &   4.5  &   0.15 &   0.6  &0.05	& 0.07	& 0.05	& 0.05	& -0.08 \\
 4256  &  5020 &   4.3  &   0.08 &   1.1  &--	&-- 	& --	&-- 	& --      \\
 4307  &  5889 &    4.0   &  -0.18 &   1.1  &--	&-- 	& --	&-- 	& --      \\
 4614  &  5965 &   4.4  &  -0.24 &   1.1  &0.18	& 0.03	& 0.19	& 0.05	& 0.03  \\
 5294  &  5779 &   4.1  &  -0.17 &   1.3  &0.19	& 0.00	& 0.11	& 0.03	& 0.26  \\
 6660  &  4759 &   4.6  &   0.08 &   1.4  &--	& --	& --	& --	& --      \\
 7590  &  5962 &   4.4  &   -0.10 &   1.4  &0.15	& 0.04	& 0.16	& 0.05	& 0.19  \\
 7924  &  5165 &   4.4  &  -0.22 &   1.1  &0.19	& 0.00	& 0.22	& 0.04	& 0.16  \\
 8648  &  5790 &   4.2  &   0.12 &   1.1  &--	& --	& 0.14	& 0.06	& -0.05 \\
 9407  &  5666 &   4.45 &   0.05 &   0.8  &0.09	& 0.06	& -0.02	& 0.03	& 0.12  \\
 9826  &  6074 &    4.0   &   0.10  &   1.3  &-0.03& 0.00	& -0.10	& 0.06	& -0.08 \\
10086  &  5696 &   4.3  &   0.13 &   1.2  &--	& --	& --	&-- 	& --    \\
10307  &  5881 &   4.3  &   0.02 &   1.1  &--	& --	& --	&-- 	& -0.10 \\
10476  &  5242 &   4.3  &  -0.05 &   1.1  &--	& -- 	& --	& --	& -0.13 \\
10780  &  5407 &   4.3  &   0.04 &   0.9  &0.16	& 0.06	& -0.02	& 0.03	& -0.12 \\
11007  &  5980 &    4.0   &   -0.20 &   1.1  &0.13	& 0.05	& 0.15	& 0.06	& 0.29  \\
11373  &  4783 &   4.65 &   0.08 &    1.0   &--	& --	& --	& --	& --      \\
12846  &  5766 &   4.5  &  -0.24 &   1.2  &0.29	& 0.04	& 0.19	& 0.07	& 0.33  \\
13507  &  5714 &   4.5  &  -0.02 &   1.1  &0.09	& --	& 0.12	& 0.05	& -0.01 \\
14374  &  5449 &   4.3  &  -0.09 &   1.1  &0.06	& --	& 0.07	& 0.06	& 0.18  \\
16160  &  4829 &   4.6  &  -0.16 &   1.1  &0.21	& --	& 0.21	& 0.00	& 0.25  \\
17674  &  5909 &    4.0   &  -0.14 &   1.1  &0.06	& 0.06	& 0.06	& 0.12	& 0.28  \\
17925  &  5225 &   4.3  &  -0.04 &   1.1  &--	& --	& --	& --	& --      \\
18632  &  5104 &   4.4  &   0.06 &   1.4  &--	& --	& --	& --	& --      \\
18803  &  5665 &   4.55 &   0.14 &   0.8  &0.03	& 0.07	& -0.03	& 0.05	& -0.12 \\
19019  &  6063 &    4.0   &  -0.17 &   1.1  &0.22	& --	& 0.21	& 0.06	& 0.16  \\
19373  &  5963 &   4.2  &   0.06 &   1.1  &-0.06& 0.02	& -0.02	& 0.05	& 0.01  \\
20630  &  5709 &   4.5  &   0.08 &   1.1  &0.07	& 0.04	& 0.00	& 0.03	& -0.06 \\
22049  &  5084 &   4.4  &  -0.15 &   1.1  &--	& --	& --	& --	& --      \\
22484  &  6037 &   4.1  &  -0.03 &   1.1  &0.12	& 0.03	& 0.07	& 0.02	& 0.15  \\
22556  &  6155 &   4.2  &  -0.17 &   1.1  &--	& --	& --	& --	& --      \\
24053  &  5723 &   4.4  &   0.04 &   1.1  &0.06	& 0.04	& 0.05	& 0.05	& 0.13  \\
24238  &  4996 &   4.3  &  -0.46 &    1.0   &0.28	& --	& 0.31	& 0.07	& 0.23  \\
24496  &  5536 &   4.3  &  -0.13 &   1.5  &0.18	& 0.03	& 0.14	& 0.05	& 0.02  \\
25665  &  4967 &   4.7  &   0.01 &   1.2  &--	&-- 	& --	& --	& --      \\
25680  &  5843 &   4.5  &   0.05 &   1.1  &--	&-- 	& --	& --	& --      \\
26923  &  5920 &   4.4  &  -0.03 &    1.0   &0.12	& 0.06	& 0.12	& 0.05	& -0.10 \\
28005  &  5980 &   4.2  &   0.23 &   1.1  &-0.03& 0.04	& -0.06	& 0.03	& 0.14  \\
28447  &  5639 &    4.0   &  -0.09 &   1.1  &0.14	& 0.04	& 0.17	& 0.03	& 0.08  \\
29150  &  5733 &   4.3  &    0.00   &   1.1  &0.06	& 0.01	& 0.04	& 0.02	& -0.13 \\
29310  &  5852 &   4.2  &   0.08 &   1.4  &--	& --	& --	& --	& --      \\
29645  &  6009 &    4.0   &   0.14 &   1.3  &-0.07& 0.05	& -0.09	& 0.04	& -0.12 \\
30495  &  5820 &   4.4  &  -0.05 &   1.3  &--	& --	& --	& 	& --      \\
33632  &  6072 &   4.3  &  -0.24 &   1.1  &0.11	& 0.04	& 0.19	& 0.04	& 0.28  \\
34411  &  5890 &   4.2  &   0.10  &   1.1  &--	&-- 	& --	&-- 	&       \\
37008  &  5016 &   4.4  &  -0.41 &   0.8  &0.18	&-- 	& 0.26	&-- 	& 0.18  \\
37394  &  5296 &   4.5  &   0.09 &   1.1  &--	&-- 	& --	& --	& --      \\
38858  &  5776 &   4.3  &  -0.23 &   1.1  &0.19	& 0.09	& 0.18	& 0.07	& 0.12  \\
39587  &  5955 &   4.3  &  -0.03 &   1.5  &0.11	& 0.03	& 0.05	& 0.03	& 0.20  \\
40616  &  5881 &    4.0   &  -0.22 &   1.1  &0.17	& 0.04	& 0.15	& 0.03	& 0.16  \\
41330  &  5904 &   4.1  &  -0.18 &   1.2  &0.15	& 0.05	& 0.19	& 0.06	& 0.32  \\
41593  &  5312 &   4.3  &  -0.04 &   1.1  &-0.02& 0.11	& 0.01	& 0.08	& -0.14 \\
42618  &  5787 &   4.5  &  -0.07 &    1.0   &0.11	& 0.03	& 0.17	& 0.05	& 0.01  \\
42807  &  5719 &   4.4  &  -0.03 &   1.1  &0.09	& 0.08	& 0.06	& 0.05	& 0.05  \\
43587  &  5927 &   4.1  &  -0.11 &   1.3  &0.01	& 0.06	& 0.07	& 0.05	& 0.30  \\
43856  &  6143 &   4.1  &  -0.19 &   1.1  &0.19	& 0.04	& 0.09	& 0.04	& --      \\
43947  &  6001 &   4.3  &  -0.24 &   1.1  &0.06	& 0.04	& 0.12	& 0.03	& 0.33  \\
45088  &  4959 &   4.3  &  -0.21 &   1.2  &0.33	& 0.07	& 0.21	& 0.00	& 0.25  \\
47752  &  4613 &   4.6  &  -0.05 &   0.2  &--	& --	& --	& --	& -0.13 \\
48682  &  5989 &   4.1  &   0.05 &   1.3  &-0.08& 0.09	& -0.05	& 0.04	& 0.07  \\
50281  &  4712 &   3.9  &   -0.20 &   1.6  &--	& --	& --	& --	& --      \\
50692  &  5911 &   4.5  &   -0.10 &   0.9  &0.11	& 0.06	& 0.12	& 0.03	& 0.24  \\
51419  &  5746 &   4.1  &  -0.37 &   1.1  &0.32	& 0.11	& 0.25	& 0.08	& 0.39  \\
51866  &  4934 &   4.4  &    0.00   &    1.0   &--	& --	& --	& --	& --      \\
53927  &  4860 &   4.64 &  -0.22 &   1.2  &0.17	& 0.04	& 0.27	& 0.00	& 0.11  \\
54371  &  5670 &   4.2  &   0.06 &   1.2  &--	& --	& --	& --	&--       \\
55575  &  5949 &   4.3  &  -0.31 &   1.1  &0.18	& 0.04	& 0.21	& 0.06	& 0.35  \\
58595  &  5707 &   4.3  &  -0.31 &   1.2  &0.20	& 0.08	& 0.21	& 0.04	& 0.30  \\
59747  &  5126 &   4.4  &  -0.04 &   1.1  &-0.04& 0.07	& -0.01	& --	& -0.04 \\
61606  &  4956 &   4.4  &  -0.12 &   1.3  &--	& --	& --	& --	& --      \\
62613  &  5541 &   4.4  &   -0.10 &   1.1  &0.14	& 0.08	& 0.11	& 0.08	& 0.14  \\
63433  &  5693 &   4.35 &  -0.06 &   1.9  &0.08	& 0.00	& 0.23	& 0.03	& 0.15   \\
64468  &  5014 &   4.2  &    0.00   &   1.2  &0.07	& 0.00	& -0.02	& 0.06	& -0.13  \\
64815  &  5864 &    4.0   &  -0.33 &   1.1  &0.28	& 0.08	& 0.23	& 0.08	& --       \\
65874  &  5936 &    4.0   &   0.05 &   1.3  &--	& --	& --	& --	& --       \\
66573  &  5821 &   4.6  &  -0.53 &   1.1  &0.38	& 0.06	& 0.37	& 0.05	& 0.45   \\
68638  &  5430 &   4.4  &  -0.24 &   1.1  &0.14	& 0.06	& 0.09	& 0.05	& 0.23   \\
70923  &  5986 &   4.2  &   0.06 &   1.1  &0.00	& 0.05	& -0.07	& 0.05	& 0.06   \\
71148  &  5850 &   4.2  &    0.00   &   1.1  &--	& --	& --	& --	& -0.13  \\
72760  &  5349 &   4.1  &   0.01 &   1.1  &--	& --	& --	& --	& --       \\
72905  &  5884 &   4.4  &  -0.07 &   1.5  &--	& --	& --	& --	& --       \\
73344  &  6060 &   4.1  &   0.08 &   1.1  &0.02	& 0.04	& -0.11	& 0.05	& -0.06  \\
73667  &  4884 &   4.4  &  -0.58 &   0.9  &0.42	& 0.06	& 0.28	& 0.05	& 0.55   \\
75732  &  5373 &   4.3  &   0.25 &   1.1  &-0.07& 0.04	& -0.13	& 0.06	& 0.02   \\
75767  &  5823 &   4.2  &  -0.01 &   0.9  &--	& --	& --	&-- 	&--        \\
76151  &  5776 &   4.4  &   0.05 &   1.1  &--	& --	& --	&-- 	&--        \\
79969  &  4825 &   4.4  &  -0.05 &    1.0   &--	& --	& --	& --	& --       \\
82106  &  4827 &   4.1  &  -0.11 &   1.1  &0.08	& 0.07	& 0.04	& 0.04	& 0.25   \\
82443  &  5334 &   4.4  &  -0.03 &   1.3  &-0.03& 0.08	& 0.10	& 0.03	& 0.10   \\
87883  &  5015 &   4.4  &    0.00   &   1.1  &--	& --	& --	& --	& --       \\
88072  &  5778 &   4.3  &    0.00   &   1.1  &0.05	& 0.03	& 0.07	& 0.03	& 0.02   \\
89251  &  5886 &    4.0   &  -0.12 &   1.1  &--	& --	& 0.14	& 0.08	&--        \\
89269  &  5674 &   4.4  &  -0.23 &   1.1  &0.19	& 0.05	& 0.09	& 0.05	& 0.17   \\
91347  &  5931 &   4.4  &  -0.43 &   1.1  &0.24	& 0.05	& 0.26	& 0.03	& 0.50   \\
94765  &  5077 &   4.4  &  -0.01 &   1.1  &0.15	& 0.06	& 0.01	& 0.04	& -0.07  \\
95128  &  5887 &   4.3  &   0.01 &   1.1  &0.06	& 0.00	& 0.02	& 0.03	& 0.01   \\
97334  &  5869 &   4.4  &   0.06 &   1.2  &-0.02& 0.04	& -0.04	& 0.03	& 0.06   \\
97658  &  5136 &   4.5  &  -0.32 &   1.2  &0.29	& 0.07	& 0.27	& 0.04	& 0.21   \\
98630  &  6060 &    4.1   &   0.22 &   1.4  &--	& --	& --	& --	& --       \\
101177 &  5932 &   4.1  &  -0.16 &   1.1  &0.13	& 0.07	& 0.14	& 0.03	& 0.25   \\
102870 &  6055 &    4.0   &   0.13 &   1.4  &-0.09& 0.05	& -0.04	& 0.05	& 0.09   \\
105631 &  5416 &   4.4  &   0.16 &   1.2  &--	& --	& --	& --	& --       \\
107705 &  6040 &   4.2  &   0.06 &   1.4  &--	& --	& --	& --	& --       \\
108954 &  6037 &   4.4  &  -0.12 &   1.1  &--	& --	& --	& --	& 0.21   \\
109358 &  5897 &   4.2  &  -0.18 &   1.1  &0.14	& 0.05	& 0.12	& 0.03	& 0.32   \\
110463 &  4950 &   4.5  &  -0.05 &   1.2  &--	& --	& --	&-- 	& --       \\
110833 &  5075 &   4.3  &    0.00   &   1.1  &--	& --	& --	&-- 	& --       \\
111395 &  5648 &   4.6  &   0.10  &   0.9  &0.02	& 0.09	& -0.05	& 0.00	& --       \\
112758 &  5203 &   4.2  &  -0.56 &   1.1  &--	& 	& --	& --	& --       \\
114710 &  5954 &   4.3  &   0.07 &   1.1  &-0.05& 0.04	& -0.03	& 0.05	& -0.05  \\
115383 &  6012 &   4.3  &   0.11 &   1.1  &-0.06& 0.03	& 0.01	& 0.03	& 0.01   \\
115675 &  4745 &   4.45 &   0.02 &    1.0   &--	& --	& --	& --	& --       \\
116443 &  4976 &   3.9  &  -0.48 &   1.1  &--	& --	& --	& --	& 0.35   \\
116956 &  5386 &   4.55 &   0.08 &   1.2  &0.06	& 0.06	& -0.01	& 0.03	& -0.21  \\
117043 &  5610 &   4.5  &   0.21 &   0.4  &-0.02& 0.03	& -0.14	& 0.03	& -0.09  \\
119802 &  4763 &    4.0   &  -0.05 &   1.1  &--	&-- 	& --	& --	& --       \\
122064 &  4937 &   4.5  &   0.07 &   1.1  &--	&-- 	& --	& --	& -0.05  \\
124642 &  4722 &   4.65 &   0.02 &   1.3  &--	& --	& --	& --	&--        \\
125184 &  5695 &   4.3  &   0.31 &   0.7  &--	& --	& --	& --	&--        \\
126053 &  5728 &   4.2  &  -0.32 &   1.1  &0.19	& 0.05	& 0.16	& 0.05	& 0.46   \\
127506 &  4542 &   4.6  &  -0.08 &   1.2  &--	& --	& --	&-- 	& --       \\
128311 &  4960 &   4.4  &   0.03 &   1.3  &0.16	& 0.05	& 0.00	& 0.03	& --       \\
130307 &  4990 &   4.3  &  -0.25 &   1.4  &0.27	& 0.05	& 0.23	& 0.03	& --       \\
130948 &  5943 &   4.4  &  -0.05 &   1.3  &--	& --	& --	& --	& --       \\
131977 &  4683 &   3.7  &  -0.24 &   1.8  &--	& --	& --	& --	&--        \\
135599 &  5257 &   4.3  &  -0.12 &    1.0   &0.07	& 0.04	& 0.11	& 0.05	& 0.21   \\
137107 &  6037 &   4.3  &    0.00   &   1.1  &0.05	& 0.03	& 0.05	& 0.04	& -0.13  \\
139777 &  5771 &   4.4  &   0.01 &   1.3  &0.11	& 0.00	& 0.08	& 0.05	& --       \\
139813 &  5408 &   4.5  &    0.00   &   1.2  &--	& --	& --	& --	& --       \\
140538 &  5675 &   4.5  &   0.02 &   0.9  &--	& --	& --	& --	& --       \\
141004 &  5884 &   4.1  &  -0.02 &   1.1  &0.12	& 0.03	& 0.08	& 0.06	& 0.04   \\
141272 &  5311 &   4.4  &  -0.06 &   1.3  &0.16	& 0.03	& 0.09	& 0.03	& 0.00   \\
142267 &  5856 &   4.5  &  -0.37 &   1.1  &--	& --	& --	& --	& --       \\
144287 &  5414 &   4.5  &  -0.15 &   1.1  &0.22	& 0.05	& 0.21	& 0.03	& 0.19   \\
145675 &  5406 &   4.5  &   0.32 &   1.1  &-0.05& 0.05	& -0.10	& 0.03	& 0.10   \\
146233 &  5799 &   4.4  &   0.01 &   1.1  &0.14	& 0.03	& 0.13	& 0.02	& -0.09  \\
149661 &  5294 &   4.5  &  -0.04 &   1.1  &0.16	& 0.07	& 0.09	& 0.05	& -0.14  \\
149806 &  5352 &   4.55 &   0.25 &   0.4  &--	&-- 	& --	&-- 	&--        \\
151541 &  5368 &   4.2  &  -0.22 &   1.3  &--	&-- 	& --	&-- 	&--        \\
153525 &  4810 &   4.7  &  -0.04 &    1.0   &--	& --	& --	& --	& --       \\
154345 &  5503 &   4.3  &  -0.21 &   1.3  &0.18	& 0.05	& 0.11	& 0.04	& 0.35   \\
156668 &  4850 &   4.2  &  -0.07 &   1.2  &0.17	& 0.04	& 0.22	& 0.07	& 0.16   \\
156985 &  4790 &   4.6  &  -0.18 &    1.0   &--	& 	& --	& 	& 0.17   \\
158633 &  5290 &   4.2  &  -0.49 &   1.3  &0.20	& 0.08	& 0.19	& 0.04	& 0.51   \\
160346 &  4983 &   4.3  &   -0.10 &   1.1  &--	& --	& --	& --	& --       \\
161098 &  5617 &   4.3  &  -0.27 &   1.1  &--	& --	& --	& --	& --        \\
164922 &  5392 &   4.3  &   0.04 &   1.1  &--	& --	& --	& --	& --        \\
165173 &  5505 &   4.3  &  -0.05 &   1.1  &0.15	& 0.06	& 0.09	&-- 	& -0.03   \\
165341 &  5314 &   4.3  &  -0.08 &   1.1  &--	& --	& --	&-- 	& -0.03   \\
165476 &  5845 &   4.1  &  -0.06 &   1.1  &--	& --	& --	&-- 	& --        \\
165670 &  6178 &    4.0   &   -0.10 &   1.5  &--	& --	& --	& --	& --        \\
165908 &  5925 &   4.1  &   -0.60 &   1.1  &0.25	& 0.04	& 0.30	& 0.06	& --0.37    \\
166620 &  5035 &    4.0   &  -0.22 &    1.0   &--	&-- 	& --	& --	& --        \\
171314 &  4608 &   4.65 &   0.07 &    1.0   &--	&-- 	& --	& --	& --        \\
174080 &  4764 &   4.55 &   0.04 &    1.0   &--	&-- 	& --	& --	& 0.08    \\
175742 &  5030 &   4.5  &  -0.03 &    2.0   &--	&-- 	& --	& --	& --        \\
176377 &  5901 &   4.4  &  -0.17 &   1.3  &0.14	& 0.07	& 0.18	& 0.02	& 0.11    \\
176841 &  5841 &   4.3  &   0.23 &   1.1  &--	&-- 	& --	&-- 	& -0.11   \\
178428 &  5695 &   4.4  &   0.14 &    1.0   &--	&-- 	& --	&-- 	& -0.17   \\
180161 &  5473 &   4.5  &   0.18 &   1.1  &--	&-- 	& --	&-- 	& --        \\
182488 &  5435 &   4.4  &   0.07 &   1.1  &--	& --	& --	&-- 	& 0.05    \\
183341 &  5911 &   4.3  &  -0.01 &   1.3  &--	& --	& --	& --	& --        \\
184385 &  5536 &   4.45 &   0.12 &   0.9  &0.00	& 0.05	& -0.03	& 0.05	& --      \\
185144 &  5271 &   4.2  &  -0.33 &   1.1  &0.10	& 0.05	& 0.06	& 0.03	& 0.37    \\
185414 &  5818 &   4.3  &  -0.04 &   1.1  &0.04	& 0.04	& 0.08	& 0.02	& -0.14   \\
186408 &  5803 &   4.2  &   0.09 &   1.1  &--	& --	& --	& --	& -0.22   \\
186427 &  5752 &   4.2  &   0.02 &   1.1  &--	& --	& --	& --	& -0.10   \\
187897 &  5887 &   4.3  &   0.08 &   1.1  &-0.01& 0.05	& -0.02	& 0.06	& -0.01   \\
189087 &  5341 &   4.4  &  -0.12 &   1.1  &--	& --	& --	& --	& 0.01    \\
189733 &  5076 &   4.4  &  -0.03 &   1.5  &0.13	& 0.04	& 0.13	& 0.05	& --        \\
190007 &  4724 &   4.5  &   0.16 &   0.8  &--	& --	& --	& --	& --        \\
190406 &  5905 &   4.3  &   0.05 &    1.0   &--	& --	& --	& --	& 0.07    \\
190470 &  5130 &   4.3  &   0.11 &    1.0   &0.00	& 0.06	& -0.10	& 0.05	& -0.14   \\
190771 &  5766 &   4.3  &   0.13 &   1.5  &--	& --	& --	& --	& --        \\
191533 &  6167 &   3.8  &   -0.10 &   1.5  &--	& --	& --	& --	& 0.19    \\
191785 &  5205 &   4.2  &  -0.12 &   1.2  &0.12	& 0.04	& 0.08	& 0.06	& 0.26    \\
195005 &  6075 &   4.2  &  -0.06 &   1.3  &--	& --	& --	& --	& 0.08    \\
195104 &  6103 &   4.3  &  -0.19 &   1.1  &--	& --	& --	& --	& --        \\
197076 &  5821 &   4.3  &  -0.17 &   1.2  &0.09	& 0.05	& 0.11	& 0.05	& 0.29    \\
199960 &  5878 &   4.2  &   0.23 &   1.1  &--	& --	& --	& --	& --        \\
200560 &  5039 &   4.4  &   0.06 &   1.1  &--	& --	& --	& --	& --        \\
202108 &  5712 &   4.2  &  -0.21 &   1.1  &0.16	& 0.08	& 0.19	& 0.03	& 0.35    \\
202575 &  4667 &   4.6  &  -0.03 &   0.5  &--	& --	& --	& --	& 0.20     \\
203235 &  6071 &   4.1  &   0.05 &   1.3  &--	& --	& --	& --	& --         \\
205702 &  6020 &   4.2  &   0.01 &   1.1  &0.06	& 0.07	& -0.01	& 0.05	& 0.21     \\
206860 &  5927 &   4.6  &  -0.07 &   1.8  &0.09	& 0.07	& 0.10	& --	& 0.16     \\
208038 &  4982 &   4.4  &  -0.08 &    1.0   &--	& --	& --	& --	& --         \\
208313 &  5055 &   4.3  &  -0.05 &    1.0   &--	& --	& --	& --	& --         \\
208906 &  5965 &   4.2  &   -0.80 &   1.7  &0.44	& 0.13	& 0.39	& 0.05	& 0.57     \\
210667 &  5461 &   4.5  &   0.15 &   0.9  &0.12	& 0.05	& 0.03	& 0.03	& --         \\
210752 &  6014 &   4.6  &  -0.53 &   1.1  &0.40	& 	& 0.41	& 0.04	& --         \\
211472 &  5319 &   4.4  &  -0.04 &   1.1  &0.04	& 0.03	& 0.06	& 0.03	& 0.06     \\
214683 &  4747 &   4.6  &  -0.46 &   1.2  &--	& --	& --	&-- 	& 0.53     \\
216259 &  4833 &   4.6  &  -0.55 &   0.5  &--	& --	& --	&-- 	& --       \\
216520 &  5119 &   4.4  &  -0.17 &   1.4  &--	& --	& --	& --	&  --        \\
217014 &  5763 &   4.3  &   0.17 &   1.1  &-0.08& 0.08	& 0.02	& 0.05	& -0.10    \\
217813 &  5845 &   4.3  &   0.03 &   1.5  &0.04	& 0.00	& 0.02	& 0.00	&  --        \\
218868 &  5547 &   4.45 &   0.21 &   0.4  &--	&-- 	& --	& --	& -0.19    \\
219538 &  5078 &   4.5  &  -0.04 &   1.1  &--	&-- 	& --	& --	& --         \\
219623 &  5949 &   4.2  &   0.04 &   1.2  &--	& --	& --	& --	& -0.12    \\
220140 &  5144 &   4.6  &  -0.03 &   2.4  &--	&-- 	& --	& --	& --         \\
220182 &  5364 &   4.5  &  -0.03 &   1.2  &--	&-- 	& --	& --	& --         \\
220221 &  4868 &   4.5  &   0.16 &   0.5  &0.09	& 0.04	& 0.06	& 0.03	& -0.04    \\
221851 &  5184 &   4.4  &  -0.09 &    1.0   &0.16	& 0.07	& 0.12	& 0.03	&--          \\
222143 &  5823 &   4.45 &   0.15 &   1.1  &--	& --	& --	& --	&--          \\
224465 &  5745 &   4.5  &   0.08 &   0.8  &--	& --	& --	& --	& --         \\
263175 &  4734 &   4.5  &  -0.16 &   0.5  &0.16	& 0.03	& 0.14	& 0.04	& 0.13     \\
BD12063 &  4859 &   4.4  &  -0.22 &   0.6  &0.19	& 0.00	& 0.21	& 0.05	& 0.34     \\
BD124499&  4678 &   4.7  &    0.00   &   0.5  &--	& --	& --	& --	& --       \\
\hline                                             	          	
thick disc&  &  &  &  &  &  &  &  \\    
\hline                                  
 245   &  5400  &  3.4  &  -0.84  &  0.7 &  0.38 & 0.12	& 0.47	& 0.05	& 0.46     \\
 3765  &  5079  &  4.3  &   0.01  &  1.1 &   0.06& 0.07	& 0.07	& 0.05	& -0.01    \\
 5351  &  4378  &  4.6  &  -0.21  &  0.5 &   --	& --	& --	& --	& 0.33     \\
 6582  &  5350  &  4.5  &  -0.83  &  0.4 &  0.28& 0.06	& 0.22	& 0.06	& 0.60     \\
13783  &  5350  &  4.1  &  -0.75  &  1.1 &  --	& --	& --	& --	& 0.57     \\
18757  &  5741  &  4.3  &  -0.25  &   1.0  &  0.15& 0.03	& 0.16	& 0.05	& 0.32     \\
22879  &  5825  &  4.42 &  -0.91  &  0.9 &  0.41& 0.03	& 0.38	& 0.03	& 0.58     \\
65583  &  5373  &  4.6  &  -0.67  &  0.7 &  0.37& 0.06	& 0.34	& 0.08	& 0.49      \\
76932  &  5840  &   4.0   &  -0.95  &   1.0  &  --	& --	& --	& --	& 0.57      \\
106516 &  6165  &  4.4  &  -0.72  &  1.1 &  --	& --	& --	& --	& --        \\
110897 &  5925  &  4.2  &  -0.45  &  1.1 &  0.15& 0.06	& 0.16	& 0.05	& 0.42      \\
135204 &  5413  &   4.0   &  -0.16  &  1.1 &  --	& --	& --	& 	& 0.03      \\
152391 &  5495  &  4.3  &  -0.08  &  1.3 &  0.13& 0.06	& 0.07	& 0.05	& 0.10      \\
157089 &  5785  &   4.0   &  -0.56  &   1.0  &  --	& --	& --	& --	& 0.43      \\
157214 &  5820  &  4.5  &  -0.29  &   1.0  &  --	& --	& --	& --	& 0.21      \\
159062 &  5414  &  4.3  &   -0.40  &   1.0  &  0.34& 0.03	& 0.23	& 0.05	& 0.27      \\
165401 &  5877  &  4.3  &  -0.36  &  1.1 &  --	& --	& --	& --	& 0.13      \\
190360 &  5606  &  4.4  &   0.12  &  1.1 &  -0.02& 0.03	& 0.04	& --	& 0.15      \\
201889 &  5600  &  4.1  &  -0.85  &  1.2 &  0.40& 0.03	& 0.41	& 0.05	& 0.57      \\
201891 &  5850  &  4.4  &  -0.96  &   1.0  &  0.45& 0.03	& 0.37	& 0.03	& 0.58      \\
204521 &  5809  &  4.6  &  -0.66  &  1.1 &  0.40& 0.03	& 0.36	& 0.05	& 0.58      \\
\hline                                    
Hercules stream   &  &  &  &  &  &  &  & \\
\hline                                                                     
13403   & 5724  &   4.0  &   -0.31  &  1.1&  0.20	& 0.03	& 0.19	& 0.05	& 0.48      \\   
19308   & 5844  &  4.3 &    0.08  &  1.1&  -0.03& 0.03	& -0.04	& 0.02	& 0.14      \\   
23050   & 5929  &  4.4 &   -0.36  &  1.1&  0.23	& 0.07	& 0.22	& 0.02	& 0.43      \\   
30562   & 5859  &   4.0  &    0.18  &  1.1&  --	& --      & --	& --	& --          \\ 
64606   & 5250  &  4.2 &   -0.91  &  0.8&  0.33	& 0.05	& 0.35	& 0.05	& 0.73      \\   
68017   & 5651  &  4.2 &   -0.42  &  1.1&  --	& --	& --	& --	& --          \\ 
81809   & 5782  &   4  &   -0.28  &  1.3&  0.18	& 0.03	& 0.15	& 0.03	& 0.25      \\   
107213  & 6156  &  4.1 &    0.07  &  1.6&  -0.05& 0.07	& -0.10	& 0.04	& 0.15      \\   
139323  & 5204  &  4.6 &    0.19  &  0.7&  -0.09& 0.06	& 0.04	& 0.03	& -0.02     \\   
139341  & 5242  &  4.6 &    0.21  &  0.9&  --	& --	& --	& --	& --          \\ 
144579  & 5294  &  4.1 &    -0.70  &  1.3&  0.35	& 0.03	& 0.35	& 0.04	& 0.67      \\   
159222  & 5834  &  4.3 &    0.06  &  1.2&  0.01	& 0.05	& 0.00	& 0.06	& -0.04     \\   
159909  & 5749  &  4.1 &    0.06  &  1.1&  --	& --	& --	& --	& --          \\ 
215704  & 5418  &  4.2 &    0.07  &  1.1&  --	& --	& --	& --	& --          \\ 
218209  & 5705  &  4.5 &   -0.43  &   1.0 &  0.32	& 0.03	& 0.32	& 0.05	& 0.50      \\   
221354  & 5242  &  4.1 &   -0.06  &  1.2&  0.06	& 	& 0.09	& 	& 0.08      \\   
\hline                                      
nonclassified   &  &  &  &  &  &  &  &   \\                                            
\hline                                                                       
 4628  &  4905 &   4.6  &  -0.36 &   0.5 & --	& --    & 0.24	& 0.05	& --          \\
 4635  &  5103 &   4.4  &   0.07 &   0.8 & 0.05	& 0.10	& 0.11	& 0.10	& 0.15       \\
10145  &  5673 &   4.4  &  -0.01 &   1.1 & --	& --	& --	& --	& --           \\
12051  &  5458 &   4.55 &   0.24 &   0.5 & --	& --	& --	& --	& --           \\
13974  &  5590 &   3.8  &  -0.49 &   1.1 & 0.14	& 0.03	& 0.07	& 0.03	& 0.31       \\
17660  &  4713 &   4.75 &   0.17 &   1.3 & --	& --	& --	& --	& --           \\
20165  &  5145 &   4.4  &  -0.08 &   1.1 & --	& --	& --	& --	& --           \\
24206  &  5633 &   4.5  &  -0.08 &   1.1 & 0.10	& 0.00	& 0.12	& 0.05	& 0.20       \\
32147  &  4945 &   4.4  &   0.13 &   1.1 & 0.02	& 0.04	& -0.01	& 0.03	& 0.29       \\
45067  &  6058 &    4.0   &  -0.02 &   1.2 & --	& --	& --	& --	&--            \\
84035  &  4808 &   4.8  &   0.25 &   0.5 & --	& --	& --	& --	&--            \\
86728  &  5725 &   4.3  &   0.22 &   0.9 & --	& --	& --	& --	& --           \\
90875  &  4788 &   4.5  &   0.24 &   0.5 & --	& --	& --	& --	& --           \\
117176 &  5611 &    4.0   &  -0.03 &    1.0  & 0.10	& 0.05	& 0.11	& 0.03	& 0.25       \\
117635 &  5230 &   4.3  &  -0.46 &   0.7 & --	& --	& --	& --	& --           \\
154931 &  5910 &    4.0   &   -0.10 &   1.1 & --	& --	& --	& --	& --           \\
159482 &  5620 &   4.1  &  -0.89 &    1.0  & --	&-- 	& --	& --	& --           \\
168009 &  5826 &   4.1  &  -0.01 &   1.1 & --	&-- 	& --	& --	& --           \\
173701 &  5423 &   4.4  &   0.18 &   1.1 & -0.01& 0.07	& -0.02	& 0.05	& 0.14       \\
182736 &  5430 &   3.7  &  -0.06 &    1.0  & 0.10	& 0.03	& 0.11	& 0.04	& 0.23       \\
184499 &  5750 &    4.0   &  -0.64 &   1.5 & 0.31	& 0.00	& 0.33	& 0.03	& 0.66       \\
184768 &  5713 &   4.2  &  -0.07 &   1.1 & --	&-- 	& --	& --	&  --          \\
186104 &  5753 &   4.2  &   0.05 &   1.1 & --	&-- 	& --	& --	&  --          \\
215065 &  5726 &    4.0   &  -0.43 &   1.1 & 0.27	& 0.03	& 0.18	& 0.04	& 0.35       \\
219134 &  4900 &   4.2  &   0.05 &   0.8 & --	& --	& --	& --	&  --          \\
219396 &  5733 &    4.0   &   -0.10 &   1.2 & 0.09	& 0.03	& 0.15	& 0.05	& 0.32       \\
224930 &  5300 &   4.1  &  -0.91 &   0.7 & 0.33	& 0.05	& 0.20	& 0.02	& 0.61       \\
\hline                                    
         
\end{longtable}                             
                                         
\label{lastpage}                            
                                           
\bsp                                        
                                           
\end{document}